\documentclass[fleqn,usenatbib]{mnras}

\usepackage{newtxtext,newtxmath}

\usepackage[T1]{fontenc}

\DeclareRobustCommand{\VAN}[3]{#2}
\let\VANthebibliography\thebibliography
\def\thebibliography{\DeclareRobustCommand{\VAN}[3]{##3}\VANthebibliography}

\usepackage{graphicx}	
\usepackage{amsmath}	
\usepackage{xspace}
\usepackage{ulem}
\usepackage{pdflscape}
\makeatletter 
  \patchcmd{\NAT@citex}
    {\@citea\NAT@hyper@{%
      \NAT@nmfmt{\NAT@nm}%
      \hyper@natlinkbreak{\NAT@aysep\NAT@spacechar}{\@citeb\@extra@b@citeb}%
      \NAT@date}}
    {\@citea\NAT@nmfmt{\NAT@nm}%
    \NAT@aysep\NAT@spacechar\NAT@hyper@{\NAT@date}}{}{}

  \patchcmd{\NAT@citex}
    {\@citea\NAT@hyper@{%
      \NAT@nmfmt{\NAT@nm}%
      \hyper@natlinkbreak{\NAT@spacechar\NAT@@open\if*#1*\else#1\NAT@spacechar\fi}%
        {\@citeb\@extra@b@citeb}%
      \NAT@date}}
    {\@citea\NAT@nmfmt{\NAT@nm}%
    \NAT@spacechar\NAT@@open\if*#1*\else#1\NAT@spacechar\fi\NAT@hyper@{\NAT@date}}
    {}{}
\makeatother

\newcommand{\ie}{i.e.\xspace}

\newcommand{\arepo}{\textsc{arepo}\xspace}
\newcommand{\areport}{\textsc{arepo-rt}\xspace}
\newcommand{\skirt}{\textsc{skirt}\xspace}

\newcommand{\oii}{[\ion{O}{II}]\xspace}
\newcommand{\oi}{[\ion{O}{I}]\xspace}

\newcommand{\hbeta}{\ion{H}{$\beta$}\xspace}
\newcommand{\oiii}{[\ion{O}{III}]\xspace}
\newcommand{\oire}{[\ion{O}{III}]$_{88}$\xspace}
\newcommand{\oirf}{[\ion{O}{III}]$_{52}$\xspace}
\newcommand{\halpha}{\ion{H}{$\alpha$}\xspace}
\newcommand{\nii}{[\ion{N}{II}]\xspace}
\newcommand{\lya}{\ion{Ly}{$\alpha$}\xspace}

\newcommand{\kt}{$k_\mathrm{transition}$\xspace}
\newcommand{\sii}{[\ion{S}{II}]\xspace}
\newcommand{\cii}{[\ion{C}{II}]\xspace}
\newcommand{\hii}{\ion{H}{II}\xspace}
\newcommand{\lsfr}{$L$--$\mathrm{SFR}$\xspace}
\newcommand{\thesan}{\textsc{thesan}\xspace}
\newcommand{\thesanone}{\textsc{thesan-1}\xspace}
\newcommand{\thesantwo}{\textsc{thesan-2}\xspace}
\newcommand{\thesanwc}{\textsc{thesan-wc-2}\xspace}
\newcommand{\thesanlow}{\textsc{thesan-low-2}\xspace}
\newcommand{\thesanhigh}{\textsc{thesan-high-2}\xspace}

\newcommand{\thesansdao}{\textsc{thesan-sdao-2}\xspace}

\title[Line intensity mapping in the EoR]{The \thesan project: predictions for multi-tracer line intensity mapping in the epoch of reionization}

\author[Kannan et al.]{
Rahul Kannan,$^{1}$\thanks{E-mail: \href{mailto:rahul.kannan@cfa.harvard.edu}{rahul.kannan@cfa.harvard.edu}}
Aaron Smith,$^{2}$\thanks{NHFP Einstein Fellow.}
Enrico Garaldi,$^{3}$
Xuejian Shen,$^4$
Mark Vogelsberger,$^{2}$
\newauthor~R\"udiger Pakmor,$^{3}$
Volker Springel$^{3}$
and Lars Hernquist$^{1}$
\\
$^{1}$Center for Astrophysics $\vert$ Harvard $\&$ Smithsonian, 60 Garden Street, Cambridge, MA 02138, USA\\
$^{2}$Department of Physics, Massachusetts Institute of Technology, Cambridge, MA 02139, USA\\
$^{3}$Max-Planck Institute for Astrophysics, Karl-Schwarzschild-Str.~1, D-85741 Garching, Germany\\
 $^{4}$TAPIR, California Institute of Technology, Pasadena, CA 91125, USA
}
\date{Accepted XXX. Received YYY; in original form ZZZ}

\pubyear{2021}

\begin{document}

\label{firstpage}
\pagerange{\pageref{firstpage}--\pageref{lastpage}}
\maketitle

\begin{abstract}
Line intensity mapping (LIM) is rapidly emerging as a powerful technique to study galaxy formation and cosmology in the high-redshift Universe. We present LIM estimates of select spectral lines originating from the interstellar medium (ISM) of galaxies and 21\,cm emission from neutral hydrogen gas in the Universe using the large volume, high resolution \thesan reionization simulations. A combination of sub-resolution photo-ionization modelling for \hii regions and Monte Carlo radiative transfer calculations is employed to estimate the dust-attenuated spectral energy distributions (SEDs) of high-redshift galaxies ($z\gtrsim 5.5$). We show that the derived photometric properties such as the ultraviolet (UV) luminosity function and the UV continuum slopes match observationally inferred values, demonstrating the accuracy of the SED modelling. We provide fits to the luminosity--star formation rate relation (\lsfr) for the brightest emission lines and find that important differences exist between the derived scaling relations and the widely used low-$z$ ones because the interstellar medium of reionization era galaxies is generally less metal-enriched than in their low redshift counterparts. We use these relations to construct line intensity maps of nebular emission lines and cross correlate with the 21\,cm emission. Interestingly, the wavenumber at which the correlation switches sign (\kt) depends heavily on the reionization model and to a lesser extent on the targeted emission line, which is consistent with the picture that \kt probes the typical sizes of ionized regions. The derived scaling relations and intensity maps represent a timely state-of-the-art framework for forecasting and interpreting results from current and upcoming LIM experiments. 
\end{abstract}




\section{Introduction}
The emission lines present in the spectral energy distributions (SEDs) of galaxies contain important information about the physical processes that shape them. They can be used to constrain the star-formation rate (SFR), metal and dust content, nature of ionizing radiation sources, and the temperature and density structure of the interstellar medium (ISM). For example, recombination cascades in hydrogen atoms ionized by newly formed young stars produce line emission, including the well-known lines of Lyman~$\alpha$ ($1216$\,\AA) and the Balmer series \halpha ($6563$\,\AA) and \hbeta ($4861$\,\AA). These strong emission lines are the most traditional indicators of the presence of star formation in galaxies \citep{Kennicutt1998}. Moreover, emission line ratios between two co-spatially emitted lines can be used to infer the dust content in galaxies, e.g. the so-called Balmer decrement in the case of \halpha/\hbeta. Likewise, strong metal emission lines at optical wavelengths such as \oii ($3726, 3729$\,\AA), \oiii ($4959, 5007$\,\AA) and \nii ($6584$\,\AA) can be used to trace the SFR in addition to constraining the metal content of galaxies \citep{Kewley2013, Wuyts2016}. Molecular gas is difficult to detect due to the lack of a dipole moment in molecular hydrogen, however \ion{H}{$_2$} is traced by organic molecules so CO lines are often taken as a proxy \citep{Leroy2008}. Far-IR lines such as \oi ($63$\,\micron) and \cii ($158$\,\micron) probe partially ionized photo-dissociation regions \citep[PDRs;][]{Maiolino2015}, while the  21\,cm spin-flip transition of neutral hydrogen is used to detect the neutral phase of the ISM \citep{Ewen1951}. Additionally, combinations of emission line ratios like \nii (6584\,\AA)/\halpha, \oiii (5000\,\AA)/\hbeta, \sii ($6717, 6731$\,\AA)/H$\alpha$ and \oi ($6300$\,\AA)/H$\alpha$ are used (called BPT diagrams; \citealt{Baldwin1981}) to distinguish the ionization mechanism of nebular gas in galaxies such as normal \hii regions, planetary nebulae, objects photoionized by active galactic nuclei (AGN), and objects excited by shock-wave heating \citep{Kewley2019}. 

Measuring these emission lines in high-$z$ galaxies has only recently become possible. For example, the excellent sensitivity and resolution of the Atacama Large Millimeter/submillimeter Array (\texttt{ALMA}) has allowed us to detect some of the most distant galaxies in the Universe by using redshifted strong emission lines in the rest frame far-IR like \oiii ($88$\,\micron), \nii ($122$\,\micron) and \cii\,($158$\,\micron) \citep{Ouchi2013, Hashimoto2018, Arata2020}. The cold molecular component of the ISM has also been detected in higher J-level transitions of the CO molecule \citep{Walter2016}.  With the imminent launch of the \textit{James Webb Space Telescope} (\texttt{JWST}), we will also be able to detect rest-frame optical emission lines well into the reionization epoch. The Near-Infrared Spectrograph (NIRSpec; \citealt{B07}) aboard the \texttt{JWST} will achieve unprecedented space-based spectral sensitivity from \mbox{$0.6$\,\micron}~to \mbox{$5.3$\,\micron}, which will help detect several bright rest frame optical star-formation rate tracers, such as \halpha  and \oiii ($5007$\,\AA), out to $z\sim9.6$. This will provide accurate measurements of spectroscopic redshifts, stellar masses and ages, dust, nebular emission line properties, metallicity, and star-formation or active galactic nuclei (AGN) driven outflows.

\begin{table*}
	\centering
	\caption{A summary of the main properties of the \thesan simulations employed in this paper. From left to right the columns indicate the name of the simulation, boxsize, initial particle number, mass of the dark matter and gas particles, the (minimum) softening length of (gas) star and dark matter particles, minimum cell size at $z=5.5$, the final redshift, the escape fraction of ionizing photons from the birth cloud (if applicable) and a short description of the simulation. A complete version of this Table can be found in \citet{KannanThesan}.  }
	\label{table:simulations}
	\begin{tabular}{lccccccccc} 
		\hline
		Name & $L_\mathrm{box}$ & $N_\mathrm{particles}$ & $m_\mathrm{DM}$ & $m_\mathrm{gas}$ & $\epsilon$ & $r^\mathrm{min}_\mathrm{cell}$& $z_\mathrm{end}$ & $f_\mathrm{esc}$ & Description\\  
		& [cMpc] & & [$\mathrm{M}_\odot$] & [$\mathrm{M}_\odot$] & [ckpc] & [pc] & & &\\
		\hline
		\thesanone & $95.5$  & $2 \times 2100^3$ & $3.12 \times 10^6$ & $5.82 \times 10^5$ & $2.2$ & $\sim 10$ & $5.5$ & $0.37$ & fiducial \\
		\\
		\thesantwo & $95.5$  & $2 \times 1050^3$ & $2.49 \times 10^7$ & $4.66 \times 10^6$ & $4.1$ & $\sim 35$ & $5.5$ & $0.37$ & fiducial\\
		\thesanwc & $95.5$  & $2 \times 1050^3$ & $2.49 \times 10^7$ & $4.66 \times 10^6$ & $4.1$ & $\sim 35$ & $5.5$ & $0.43$ & weak convergence of $x_\mathrm{HI} (z)$ \\
		\thesanhigh & $95.5$  & $2 \times 1050^3$ & $2.49 \times 10^7$ & $4.66 \times 10^6$ & $4.1$ & $\sim 35$ & $5.5$ & $0.8$ & $f_{\mathrm{esc}} \propto \mathrm{M}_\mathrm{halo} (> 10^{10})$\\
		\thesanlow & $95.5$  & $2 \times 1050^3$ & $2.49 \times 10^7$ & $4.66 \times 10^6$ & $4.1$ & $\sim 35$ & $5.5$ & $0.95$ & $f_{\mathrm{esc}} \propto \mathrm{M}_\mathrm{halo} (<10^{10})$\\
		\thesansdao & $95.5$  & $2 \times 1050^3$ & $2.49 \times 10^7$ & $4.66 \times 10^6$ & $4.1$ & $\sim 35$ &  $5.5$ & 0.55 & Strong dark acoustic oscillations\\
		\hline
	\end{tabular}
\end{table*}

Another intriguing method of detecting emission lines is through line intensity mapping (LIM), which measures the spatial fluctuations in the integrated emission from spectral lines originating from many individually unresolved galaxies \citep{Visbal2010}. While galaxy surveys probe discrete objects that are bright enough to be detected, LIM is sensitive to all emission line sources in the observed volume, facilitating a complete census of the sites of emission line galaxies. Moreover, these maps can be used to trace the galaxy distribution in the same cosmological volumes as 21\,cm observations. Mapping both the galaxies themselves and the surrounding neutral gas in the IGM will dramatically improve our understanding of the interplay between the ionizing sources and the IGM throughout the EoR \citep{Lidz2011, Gong2012, Kovetz2019}.

Current and upcoming experiments aim to target a number of bright emission lines from high-$z$ galaxies. For example, the CO Mapping Array Project \citep[\texttt{COMAP-high};][]{comap} will measure the CO (2--1) line at $z\simeq5.8$--$7.8$. The Cerro Chajnantor Atacama Telescope-prime \citep[\texttt{CCAT}-p;][]{ccatp} aims to detect the \cii ($158$\,\micron) and \oiii ($88$\,\micron) lines at $z\simeq5$--$9$. The \textit{Origins} Space Telescope \citep{origins} will observe a host of far-IR lines emanating from high-$z$ galaxies including \oi ($63$\,\micron), \oiii ($52$ and $88$\,\micron), \nii ($122$\,\micron) and \cii ($158$\,\micron), to name just a few. The \textit{Spectro-Photometer for the history of the Universe, Epoch of Reionization and Ices Explorer} \citep[\texttt{SPHEREx;}][]{Dore2014} is scheduled to be launched in 2024 and is designed to measure a host of rest frame optical/UV emission lines including \halpha, \hbeta, \lya, \oii ($3726, 3729$\,\AA) and \oiii ($4959, 5007$\,\AA) over a wide redshift range ($z=0$--$12$). Finally, the proposed \textit{Cosmic Dawn Intensity Mapper} \citep[\texttt{CDIM;}][]{Cooray2019} aims to map \halpha, \lya and \oiii ($4959, 5007$\,\AA) at $z=0.2$--$10$, with a much higher sensitivity. These measurements will complement the intensity mapping estimates of the spatial fluctuations in the integrated emission of the 21\,cm spin-flip transition of the hydrogen atom from instruments such as the Precision Array for Probing the Epoch of Reionization  \citep[\texttt{PAPER};][]{PAPER}, the Low-Frequency Array \citep[\texttt{LOFAR};][]{LOFAR}, the Square Kilometer Array \citep[\texttt{SKA};][]{SKA} and the Hydrogen Epoch of Reionization Array \citep[\texttt{HERA};][]{HERA}.

One of the major drawbacks of LIM concerns foregrounds. If a foreground emission line is redshifted into the same observing band as the target line, it will be difficult to tell them apart. One method to overcome these foregrounds is to cross-correlate two different emission lines with each other. The two maps will be correlated, and their foregrounds will not, producing a cross spectrum which depends only on the two target lines \citep{Kovetz2017}. It is therefore important to model the luminosities of multiple tracers arising from high-$z$ galaxies in order to make accurate predictions for these experiments. Most current theoretical and simulation-based modelling involves using scaling relations that relate the line luminosity to the star-formation rate of the galaxy \citep{Gong2012, Gong2017, Fonseca2017, Heneka2021, Schaan2021}. However, most of these scaling relations come from observational estimates and/or theoretical calculations of emission line properties of low-$z$ galaxies. It is unclear if these scaling relations are valid in the relatively low metallicity ISM of high-$z$ galaxies. It is therefore imperative to derive self-consistent \lsfr scaling relations that capture the emission line properties of high-$z$ galaxies (see for example \citealt{Leung2020}).  In this work we use the state-of-the-art large volume, high resolution \thesan cosmological reioinization simulations to derive scaling relations for various bright emission lines present in high-$z$ galaxies. The scaling relations are used to make LIM predictions for nebular emission lines and cross-correlate them with the 21\,cm emission throughout the Epoch of Reionization in a self-consistent manner. The layout of the paper is as follows. Our methodology is introduced in Section~\ref{sec:methods}.   Main results are presented in Section~\ref{sec:results} and caveats and conclusions are given in Section~\ref{sec:conclusions}.

\section{Methods}
\label{sec:methods}
\subsection{The \thesan simulations}
\label{sec:thesan}
We use the \thesan simulations \citep{KannanThesan, GaraldiThesan, SmithThesan} to make predictions for the emission line properties of high-redshift galaxies and at the same time probe the 21\,cm emission from the IGM as reionization progresses. \thesan is a suite of of large volume ($L_\mathrm{box} = 95.5 \, \mathrm{cMpc}$) radiation-magneto-hydrodynamic simulations that self-consistently model the hydrogen reionization process and the resolved properties of the sources (galaxies and AGN) responsible for it. The simulations are performed with \areport \citep{Kannan2019} a radiation hydrodynamic extension to the moving mesh code \arepo \citep{Springel2010,Weinberger2020}. It solves the magneto-hydrodynamic (MHD) equations on an unstructured Voronoi grid constructed from a set of mesh generating points that are allowed to move along with the underlying gas flow. A quasi-Langrangian solution to the fluid equations is obtained by solving the Riemann problem  at the interfaces between moving mesh cells in the rest frame of the interface. The magnetic fields are evolved assuming ideal MHD \citep{Pakmor2013} with the eight-wave formalism outlined in \citet{Powell1999} used to control divergence errors. Gravity is solved with a Tree-PM approach that uses an oct-tree \citep{Barnes1986} algorithm to estimate the short range gravitational forces and a Particle Mesh method \citep{Gadget4} to compute the long range ones.

Radiation fields are modelled using a moment based approach that solves the zeroth and first moments of the radiative transfer equation \citep{Rybicki1986}. This gives rise to a set of hyperbolic conservation equations for photon number density and photon flux. The system of equations is closed using the M1 closure relation, that approximates the Eddington tensor based on the local properties of a cell \citep{Levermore1984}. The radiation fields are coupled to the gas via a non-equilibrium thermo-chemistry module, which self consistently calculates the ionization states and cooling rates from hydrogen and helium ($n_j \in [n_\ion{H}{I}, n_\ion{H}{II}, n_\ion{He}{I}, n_\ion{He}{II}, n_\ion{He}{III}]$), metal cooling based on the model outlined in \citet{Vogelsberger2013} and Compton cooling of the CMB (Section~3.2.1 of \citealt{Kannan2019}). Both stars and AGN act as sources of radiation. The radiation intensity and the spectral energy distribution of stars is a complex function of its age and metallicity and is taken from the Binary Population and Spectral Synthesis models \citep[BPASS;][]{BPASS2017}. The AGN radiation output is scaled linearly with the mass accretion rate with a radiation conversion efficiency of $0.2$ \citep{Weinberger2018} and a \citet{Lusso2015} parametrization for the shape of its spectrum. The ionizing photons are split into three frequency bins with the bin edges corresponding to the ionization energy of neutral hydrogen [13.6 eV, 24.6 eV), neutral helium [24.6 eV, 54.4 eV) and singly ionized helium [54.4 eV, $\infty$). The mean photoionization cross sections ($\sigma$), energy injected into the gas per interacting photon ($\mathcal{E}$), and the mean energy per photon ($e$) of each bin is assumed to be constant throughout the entire simulation and they are summarised in Table~1 of \citet{KannanThesan}.

For processes occurring on scales smaller than the resolution limit of the simulation (e.g. star formation, black hole accretion, feedback etc.) we employ the state-of-the-art IllustrisTNG \citep{Springel2018, Marinacci2018, Naiman2018, Pillepich2018b, Nelson2018} galaxy formation model. It treats the interstellar medium (ISM) as a two-phase gas where cold clumps are embedded in a smooth, hot phase produced by supernova explosions \citep{Springel2003}. Feedback from supernova explosions and stellar winds are implemented in the form of kinetic and thermal energy injection into the surrounding ISM \citep{Pillepich2018}. Metal production and the evolution of nine elements (H, He, C, N, O, Ne, Mg, Si and Fe) are tracked along with the overall gas metallicity. Black hole formation is implemented via a seeding prescription, followed by growth and feedback in two different regimes \citep[quasar- and radio-mode;][]{Weinberger2017}. Additionally, the model is augmented with a scalar dust model that tracks the production, growth, and destruction of dust using the formalism outlined in \cite{Mckinnon2016, Mckinnon2017}. Moreover, one additional parameter, $f_\mathrm{esc}$, is added to mimic the absorption of LyC photons happening below the resolution scale of the simulation. This parameter is tuned such that the simulated reionization histories approximately match the observed neutral fraction evolution in the Universe \citep[see e.g.][]{Greig2017}. We note that we only apply this escape fraction for LyC photons emitted by stars, while the AGN have an escape fraction of unity. This assumption does not affect the results because the AGN contribution to reionization is minimal ($\leq 1 \%$) in the \thesan simulations \citep{KannanThesan, Yeh2022}.

All \thesan simulations follow the evolution of a cubic patch of the universe with linear comoving size \mbox{$L_\mathrm{box} = 95.5$~cMpc}. We employ a \citet{Planck2015_cosmo} cosmology (more precisely, the one obtained from their \texttt{TT,TE,EE+lowP+lensing+BAO+JLA+H$_0$} dataset), \ie $H_0 = 100\,h\,\text{km\,s}^{-1}\text{Mpc}^{-1}$ with $h=0.6774$, $\Omega_\mathrm{m} = 0.3089$, $\Omega_\Lambda = 0.6911$, $\Omega_\mathrm{b} = 0.0486$, $\sigma_8 = 0.8159$, and $n_s = 0.9667$, where all symbols have their usual meanings. The highest resolution simulation (\thesanone) has a dark matter mass resolution of $3.12 \times 10^6 \, \text{M}_\odot$ and a baryonic mass resolution of $5.82 \times 10^6 \, \text{M}_\odot$. The gravitational forces are softened on scales of \mbox{$2.2$ ckpc} with the smallest cell sizes reaching $10$~pc. This allows us to model atomic cooling haloes throughout the entire simulation volume. The fiducial simulation is complemented with a set of medium resolution simulations designed to investigate the changes to reionization induced by numerical convergence, mass-dependent escape fractions from galaxies, alternative dark matter models, and the impact of assuming a constant radiation background in galaxy formation simulations. These additional runs employ identical initial conditions, but the mass resolution is coarser by a factor of $8$ and the softening length is increased by about a factor of two. Specifically, we incorporate the following additional runs into our LIM analysis: \thesantwo is a medium resolution run that uses the same fiducial model and escape fraction as \thesanone. \thesanwc is the same as \thesantwo but has a slightly higher escape fraction, which compensates for the lower star-formation rate in the medium resolution runs such that the total integrated number of photons emitted in \thesanone and \thesanwc runs are the same. The \thesanhigh and \thesanlow simulations use a halo mass dependent escape fraction where only haloes respectively above and below a halo mass threshold of $10^{10}\, \text{M}_\odot$ contribute to the reionization process. Finally, \thesansdao probes the impact of assuming non-standard dark matter models on the reionization process by simulating reionization with a strong Dark Acoustic Oscillation (sDAO) DM model. The properties of the simulations are all summarised in Table~\ref{table:simulations}.

\subsection{Emission line modelling}
\label{sec:emission}
The emission line luminosities of various lines depend strongly on the detailed properties of the galaxies they originate from. Galaxies with high star-formation rates will contain a large number of young massive OB stars that emit copious amounts of LyC radiation, which will interact with the surrounding ISM to produce ionized \hii regions. This photoionized and photoheated gas cools via radiative recombination processes and emission from forbidden and
fine-structure line transitions. Emission lines are also emitted from dense molecular clouds and largely neutral PDRs where the main source of heating is the far-UV radiation. As a result, the strength of these lines is sensitive to the ionization state of the gas and its metal and dust content \citep{Kennicutt1998, Kewley2019}. Therefore, an accurate characterisation of emission line luminosities requires accurate modelling of various galaxy properties such as a multiphase ISM, including resolved \hii, PDR and molecular regions, metal and dust production and distributions, and precise radiation field intensities. 

While there have been several recent attempts to model these processes in a self-consistent manner in galaxy scale simulations \citep{Katz2019, Pallottini2019, Kannan2020b, Kannan2021}, the numerical cost of these efforts force them to treat only a handful of galaxies at a time.  Larger-scale simulations that aim to simulate a representative volume of the Universe \citep{Vogelsberger2014, Schaye2015, Springel2018, VNature} do not have sufficient resolution to accurately determine the ionization and temperature structure of \hii/PDR regions. Moreover, most of these simulations use an effective equation of state description of the ISM in order to overcome the inability to resolve the multi-phase nature of the high-density gas in galaxies. Therefore, a sub-resolution model of \hii regions is required to estimate the emission line properties of galaxies from these large scale simulations. For example, \citet{Hirschmann2017, Hirschmann2019} used  \textsc{cloudy} \citep[last described in][]{cloudy} to estimate the emission line properties of simulated low redshift galaxies containing active star formation, old post-AGB stars and the presence of a central AGN. This was achieved by running a grid of physical models spanning typical ISM densities, radiation field intensities, metallicites and dust content found in the simulated galaxies and convolving them together to obtain the galaxy-wide emission. Similar methods have recently been employed to estimate the emission line properties of reionization era galaxies \citep[BlueTides and FLARES; ][]{Wilkins2020, Vijayan2021}.

 \begin{figure}
	\includegraphics[width=0.99\columnwidth]{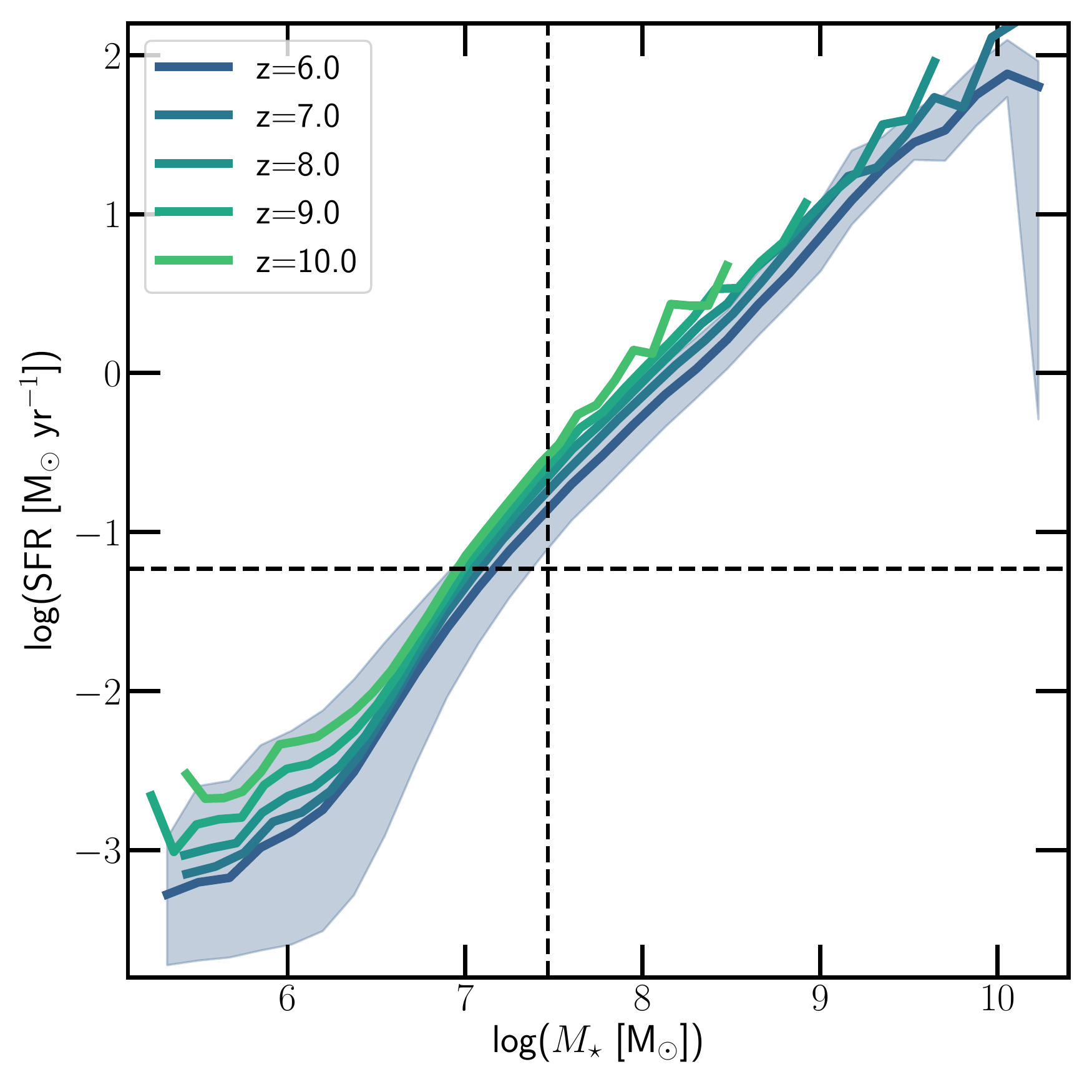}
    \caption{The star-formation rate as a function of the stellar mass of a galaxy at $z=6$--$10$ as indicated. The shaded region represents the $10^\text{th}$ and $90^\text{th}$ percentiles of the distribution at $z=6$. Similar scatter in the relation exists at other redshifts, which we have chosen not to show for the sake of clarity. For the \skirt calculations, we only consider galaxies with stellar masses at least $50$ times larger than the baryonic mass resolution of the simulation (vertical line) and with at least one star particle less than \mbox{$5$ Myr} old (horizontal line).}
    \label{fig:sfr}
\end{figure}

Of particular relevance to this paper, the IllustrisTNG simulations have been post-processed using similar schemes to estimate the high-redshift luminosity functions, dust content \citep{Vogelsberger2020}, dust attenuation curves and IR luminosity functions \citep{Shen2020, Shen2021}. We employ the same scheme as these works to estimate the emission line luminosities from the \thesan simulations. Briefly, the intrinsic nebular emission is taken from  \citet{Byler2017}, which is based on photoionization calculations using \textsc{cloudy}. The underlying stellar radiation intensities and spectra are taken from the Flexible Stellar Population Synthesis (FSPS) model \citep{Conroy2009}. The calculations assume that the fraction of escaping Lyman continuum photons is zero and nebular emission is completely determined by the gas phase metallicity and the ionization parameter. The \hii regions are assumed to be dust free. Such an assumption is necessary in order to match the low mass end of the observed high-redshift UV luminosity functions \citep{Vogelsberger2020} and the equivalent widths (EWs) of the \mbox{\oiii+\hbeta} aggregate emission line in high-$z$ galaxies \citep{Shen2020}. The gas phase metallicity of the \hii regions is chosen to be the same as the initial metallicity of the stellar particles, which is inherited from the gas cell from which a stellar particle is created. The ionization parameter, that encodes information about the intensity of the ionizing source and the geometry of the gas cloud is chosen to be $10^{-2}$ as suggested in \citet{Byler2018}. Only young stellar populations (with ages less than 10 Myrs old) are assumed to be surrounded by their birth cloud and hence the nebular emission is only active for these stars. Older stars generally escape the birth clouds and, therefore, only contribute intrinsic radiation emission as given by the FSPS model \citep{Shen2020}. A slight inconsistency arises due to the fact that the nebular emission models are computed assuming a \citet{Kroupa2001} model while the intrinsic stellar SEDs are calculated with a \citet{Chabrier2003} IMF. However, this introduces only small differences in the computed galaxy SEDs and we are confident that the inconsistency in the population synthesis models does not have a significant impact on our results for all bands and all redshifts \citep[see][for a more thorough discussion.]{Vogelsberger2020}. 

\begin{figure}
	\includegraphics[width=0.99\columnwidth]{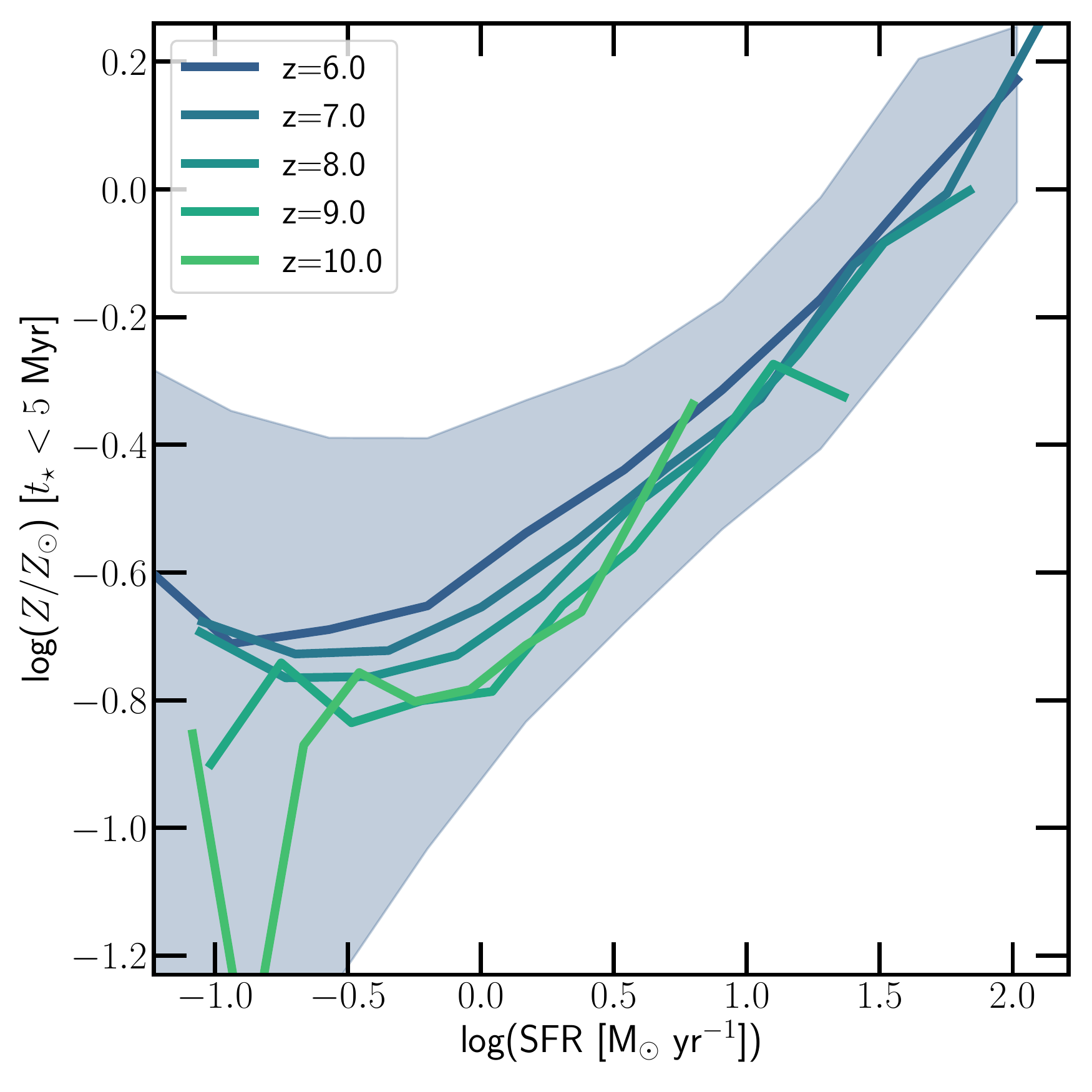}
    \caption{The star-formation rate--metallicity (in units of $Z_\odot$) relation for stars younger than \mbox{$5$ Myr} old at $z=6$--$10$ as indicated. The shaded region shows the $10^\text{th}$ and $90^\text{th}$ percentiles of the distribution at $z=6$. As the star-formation rate of the halo increases, the metallicity of newly formed stars increases considerably (by almost an order of magnitude). In contrast the redshift evolution is much less extreme.}
    \label{fig:met}
\end{figure}

 \begin{figure*}
	\includegraphics[width=0.99\textwidth]{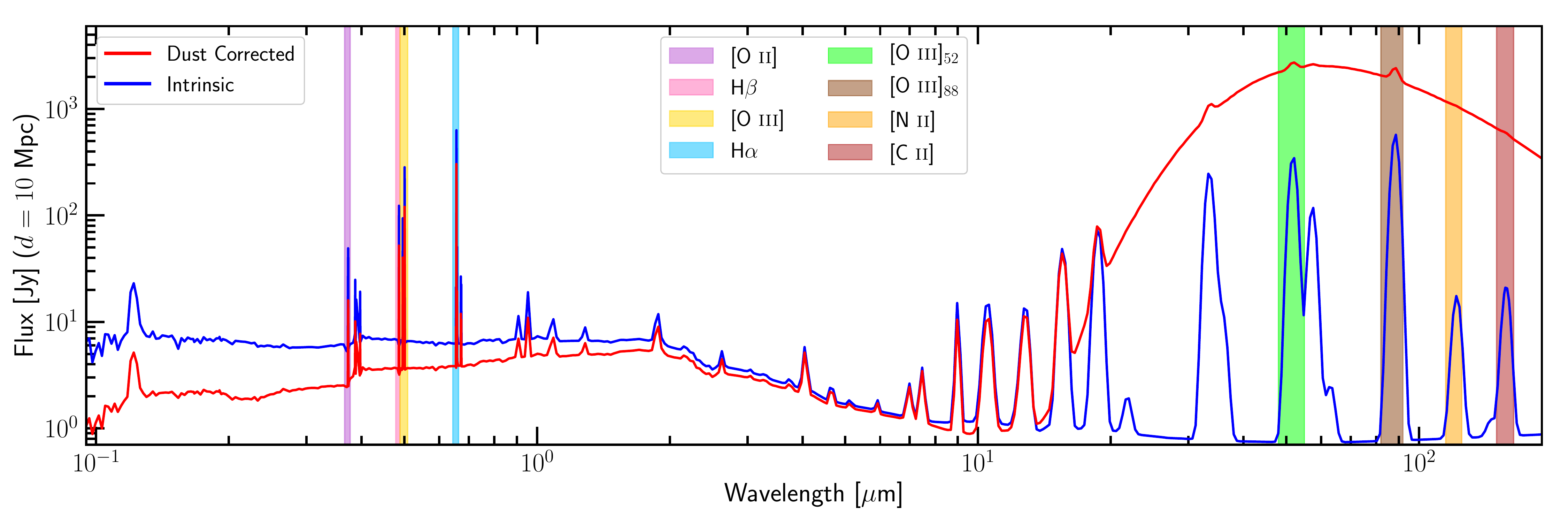}
    \caption{The rest frame spectral energy distribution of the central galaxy in the most massive halo in the \thesanone simulation \mbox{($M_\star(<2r_{\mathrm{half},\star}) \simeq 7 \times 10^{9} \mathrm{M}_\odot$)} at $z=6$. The blue curve shows the intrinsic emission and the red curve indicates the dust reprocessed SED. The different coloured vertical bands denote the positions of the various emission lines considered in this work. These are the four rest frame optical emission lines, \mbox{\oii ($3726, 3729$\,\AA)}, \mbox{\hbeta ($4861$\,\AA)}, \mbox{\oiii ($4959, 5007$\,\AA)} and \mbox{\halpha ($6563$\,\AA)}, and four fine structure emission lines in the IR, \mbox{\oiii$_{52}$ ($52$\,\micron)}, \mbox{\oiii$_{88}$ ($88$\,\micron)}, \mbox{\nii ($122$\,\micron)} and \mbox{\cii ($158$\,\micron)}.}
    \label{fig:sed}
\end{figure*}

Dust attenuation is modelled using the Monte Carlo radiative transfer code \skirt \citep[Version 8; last described in][]{skirt}, which accurately captures scattering and absorption by dust throughout the entire wavelength range considered. The radiative transfer calculations are performed on a Voronoi grid constructed from positions of the gas cells in the galaxy, mirroring the structure and geometry of the \arepo data as closely as possible. Although the \thesan simulations include self-consistent dust modelling, they predict relatively low dust-to-metal ratios (DTM) in high metallicity environments. This leads to insufficient dust attenuation producing too many high luminosity galaxies  \citep{KannanThesan}. We therefore use a spatially constant DTM that scales as $\mathrm{DTM} \simeq 0.9 \times (z/2)^{-1.92}$, where $z$ is the redshift. This relation was primarily derived to match the observed evolution of the UV luminosity function in the IllustrisTNG simulations \citep{Vogelsberger2020}. There is some evidence to suggest that the DTM ratios of high redshift galaxies are much lower than in the local Universe \citep[see for example,][]{Inoue2003, Aoyama2017, Behrens2019}. However, recent observational evidences  \citep{DeCia2016, Wiseman2017} from damped \mbox{Lyman-$\alpha$} absorbers do not show an obvious redshift dependence. In fact they show a much stronger correlation with the gas metallicity than redshift. Moreover, hydrodynamical simulations that include self-consistent dust modelling show that the DTM ratio can vary even within a single galaxy \citep{Bekki2015, Mckinnon2016}. Estimating a more sophisticated DTM relation that depends on both metallicty and redshift is beyond the scope of this work. We therefore use the relation derived in \citet{Vogelsberger2020} in this work and note that, in addition to the UV luminosity functions, this model has been relatively successful in estimating various emission line luminosities and IR luminosity functions as well \citep{Shen2020, Shen2021}.

Dust emission is modelled assuming thermal equilibrium with the local radiation field. A \citet{Draine2007} dust mixture of amorphous silicate and graphitic grains, including varying amounts of polycyclic aromatic hydrocarbons (PAHs) particles, is assumed. This model reproduces the average Milky Way extinction curve and is widely used \citep{sunrise, RR2014}. Stars act as sources of radiation, with the smoothing length of the star particles used to calculate a smoothed photon source distribution function. For each wavelength on the wavelength grid, $N_\mathrm{p}$ photon packets are launched isotropically from the smoothed positions  of the stellar particles. The photon packets then propagate through the resolved gas (dust) distribution in the ISM and interact with the dust cells randomly before they are finally collected by the photon detector, which is placed at a distance of \mbox{10 Mpc} from the galaxy centre in the $+z$ direction. We set $N_\mathrm{p}$ equal to the number of all the stellar particles bound to a galaxy with an upper and lower bound of this value set to $10^5$ and $10^2$ respectively. The wavelength grid ranges from \mbox{$0.05$\,\micron} to \mbox{$200$\,\micron}, discretised into $657$ unequal bins, with increased resolution focused around the  wavelengths of the emission lines considered in this work, which are the four rest frame optical emission lines, \mbox{\oii ($3726, 3729$\,\AA)}, \mbox{\hbeta ($4861$\,\AA)}, \mbox{\oiii ($4959, 5007$\,\AA)} and \mbox{\halpha ($6563$\,\AA)}, and four fine structure emission lines in the IR, \mbox{\oiii$_{52}$ ($52$\,\micron)}, \mbox{\oiii$_{88}$ ($88$\,\micron)}, \mbox{\nii ($122$\,\micron)} and \mbox{\cii ($158$\,\micron)}. 

Only sufficiently well-resolved, star-forming galaxies are considered, so as to ensure that the output SEDs from the \skirt calculations are reasonably converged. We consider a galaxy well resolved if the stellar mass within twice the stellar half-mass radius is greater than $50$ times the baryonic mass resolution of the simulation. This limit is indicated by the vertical dashed line in Figure~\ref{fig:sfr}, which shows the star-formation rate of a galaxy as a function of its stellar mass in \thesanone for $z=6$--$10$ as indicated. Most of the galaxies lie on the star-forming main sequence with similar slopes at all redshifts considered, although the amplitudes increase with redshift. Since most of the emission lines are related to the star-formation rate of the galaxy \citep{Kennicutt1998}, we also require that at least one star particle be less than \mbox{$5$ Myrs} old which ensures that there is recent on-going star formation within the halo. Given the resolution of the simulation, this roughly translates to haloes with a minimum star-formation rate of just below \mbox{$0.1$ M$_\odot$ yr$^{-1}$} (dashed horizontal in Figure~\ref{fig:sfr}). For all selected haloes, we compute the SEDs and emission line luminosities arising from the central \mbox{$30$ kpc} of the galactic centre. The analysis is carried out for all haloes that match these criteria at $z=6$, $z=7$, $z=8$, $z=9$ and $z=10$. This translates to $\sim 10,000$ galaxies at $z=6$ and about $\sim 20,000$ in total.

 \begin{figure*}
 \begin{center}

 		\includegraphics[width=0.495\textwidth]{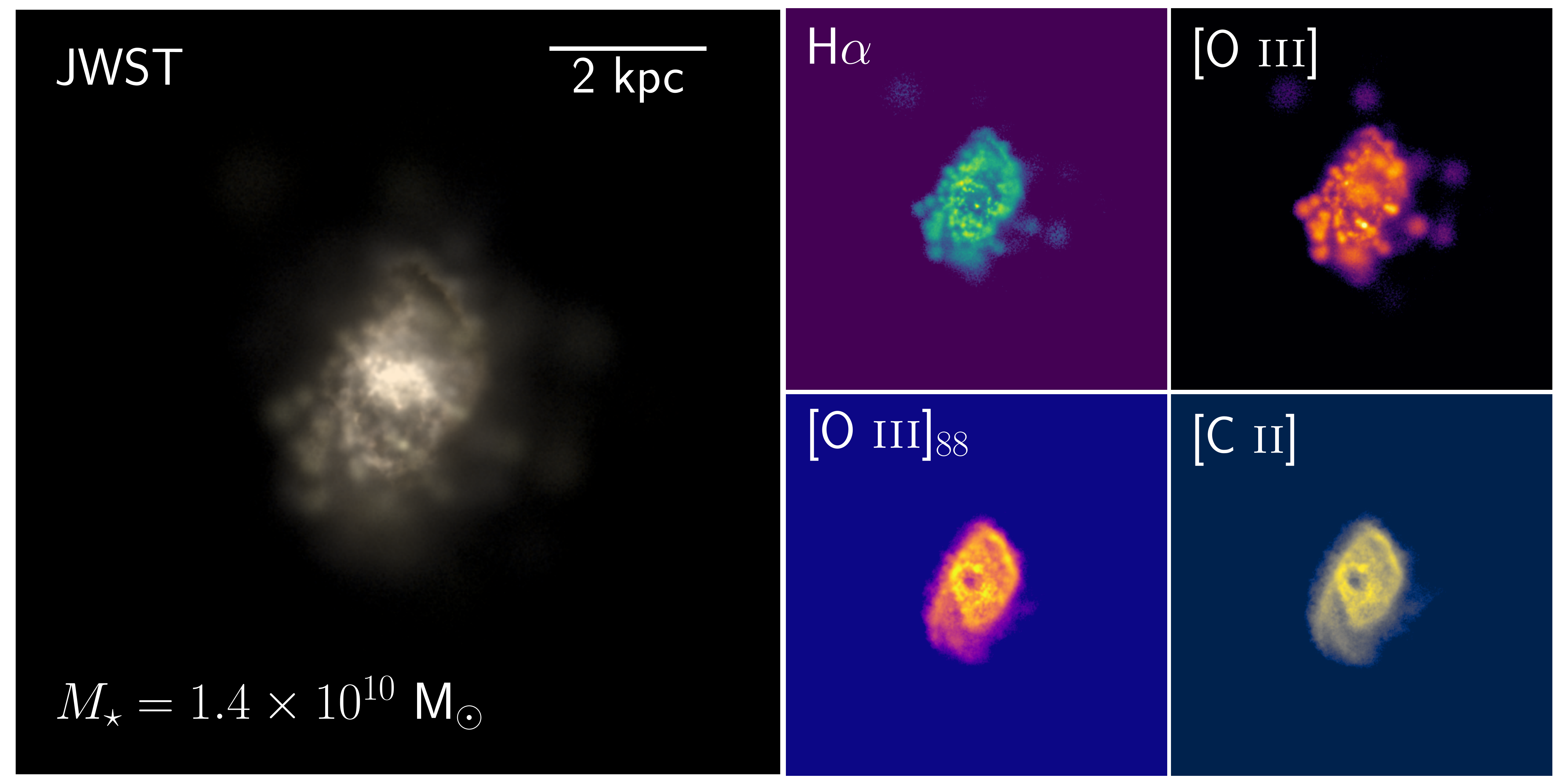}
	\includegraphics[width=0.495\textwidth]{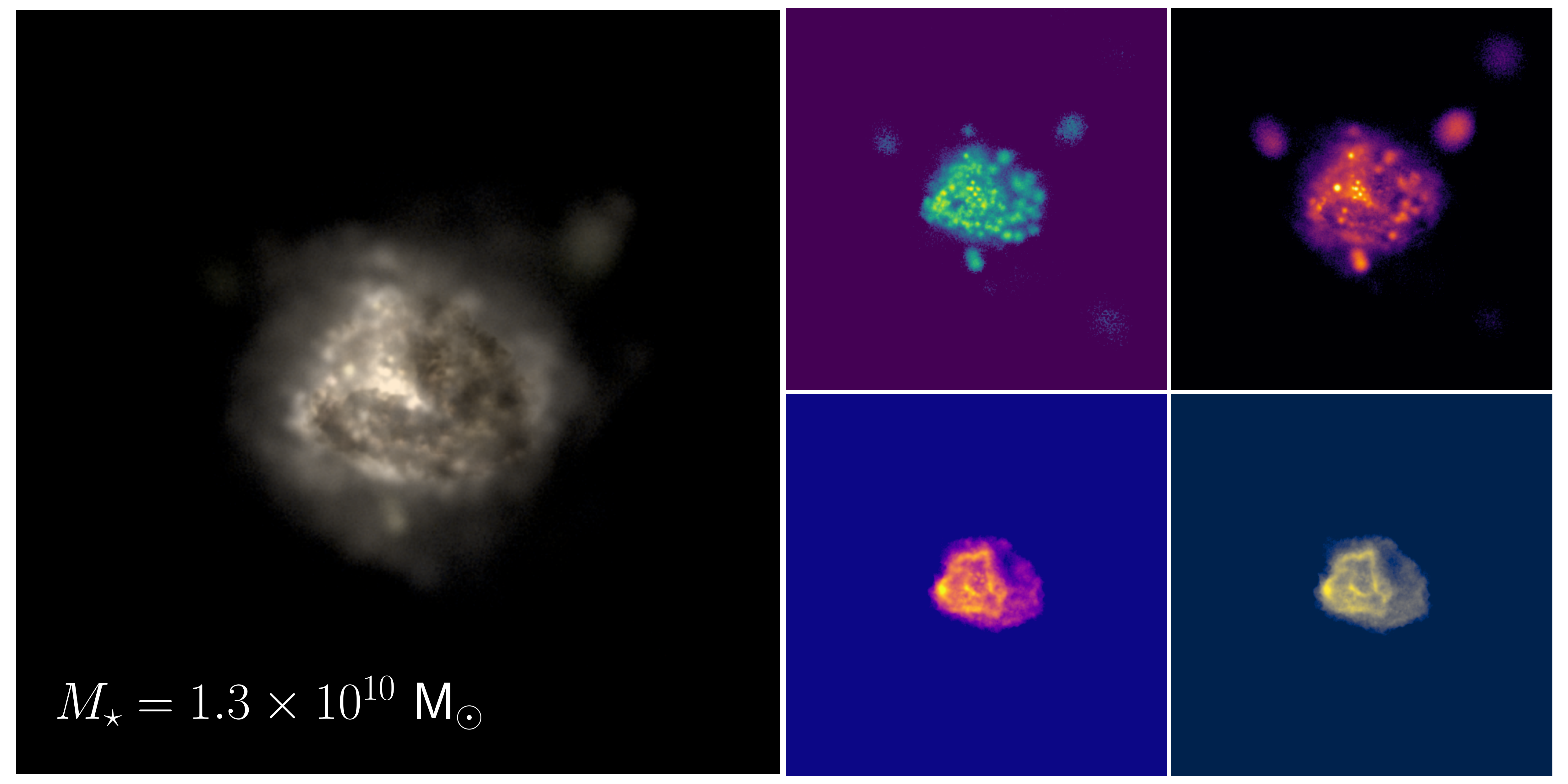}
		\includegraphics[width=0.495\textwidth]{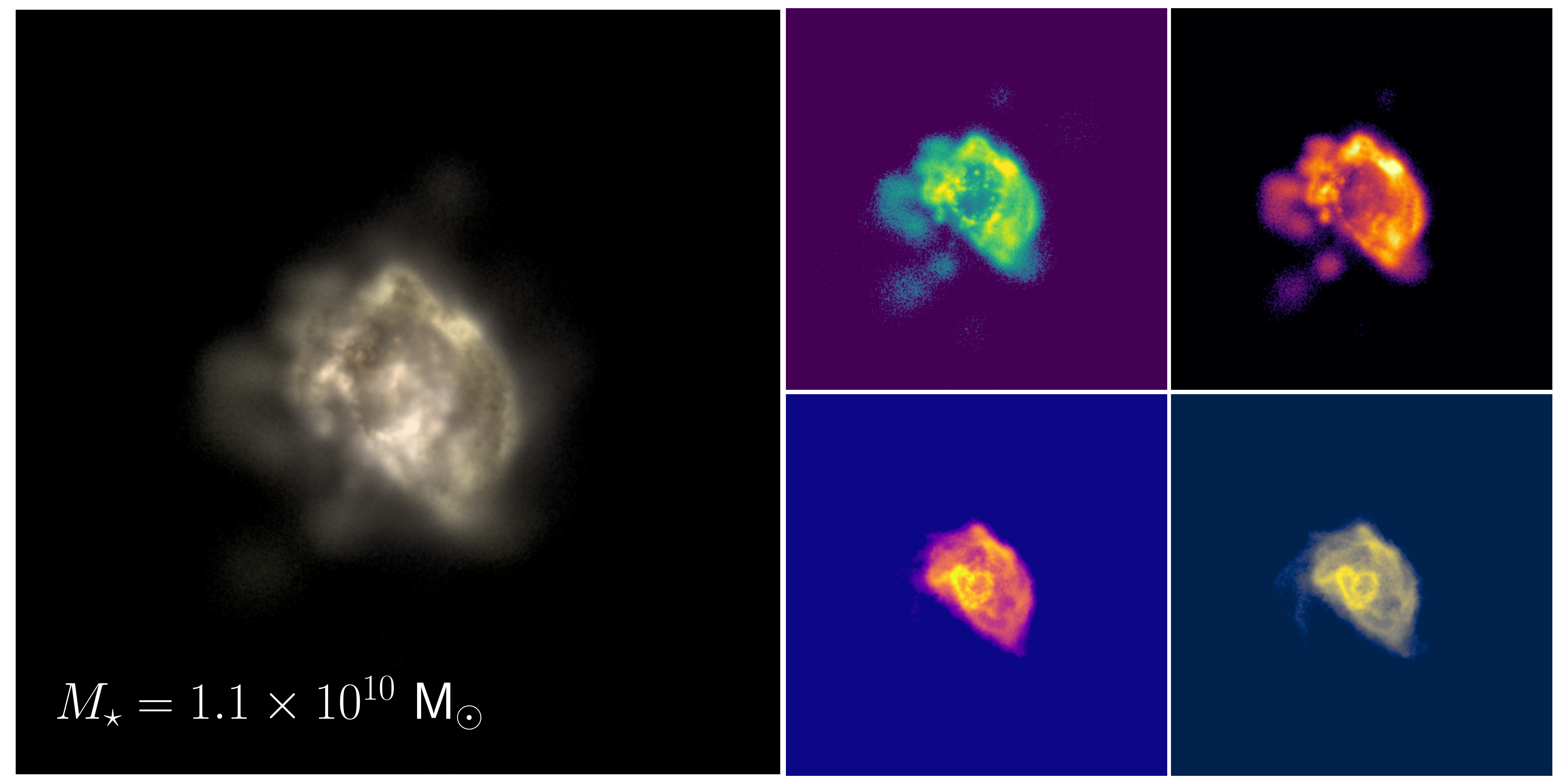}  
		\includegraphics[width=0.495\textwidth]{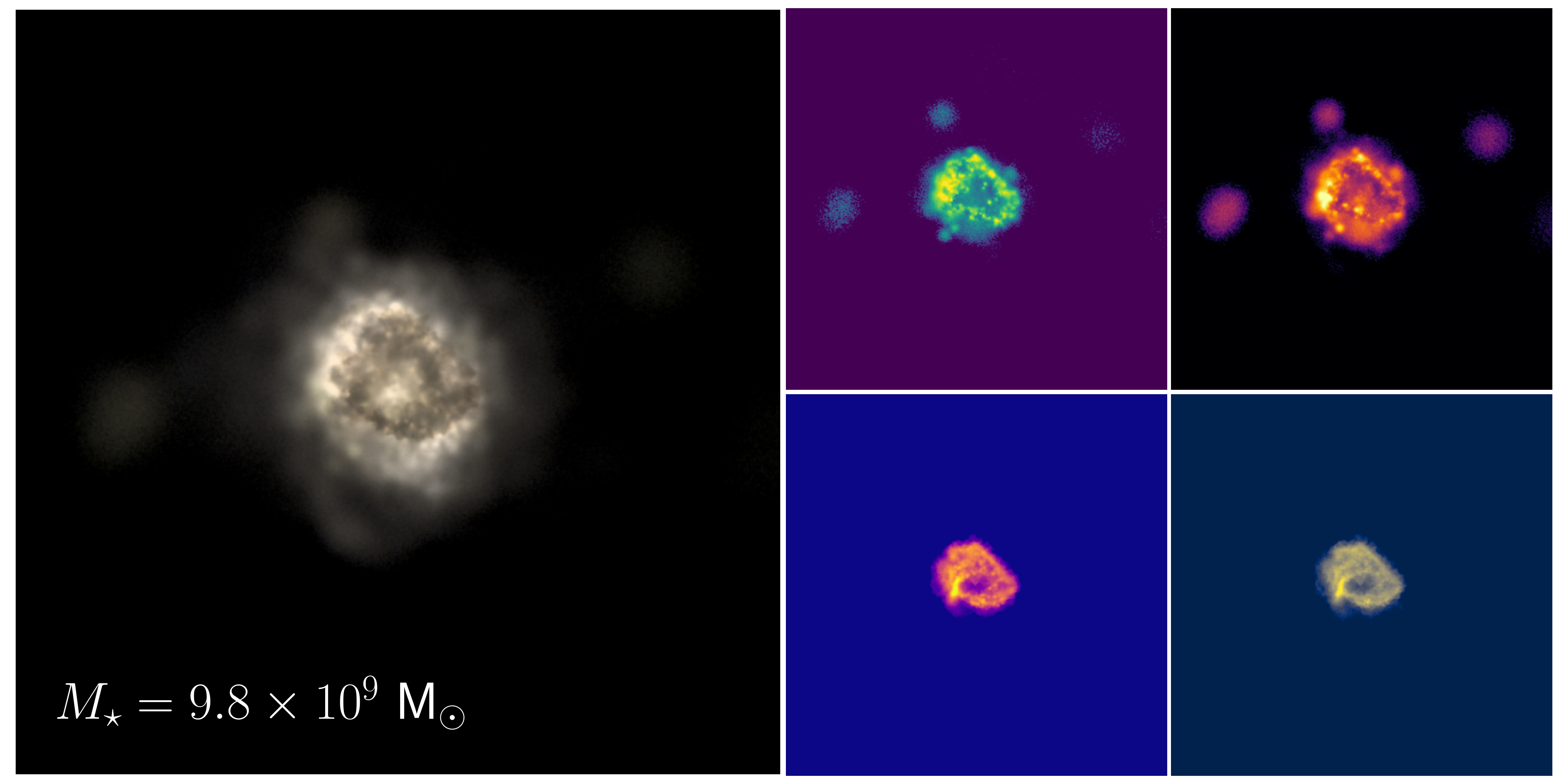}    
			\includegraphics[width=0.495\textwidth]{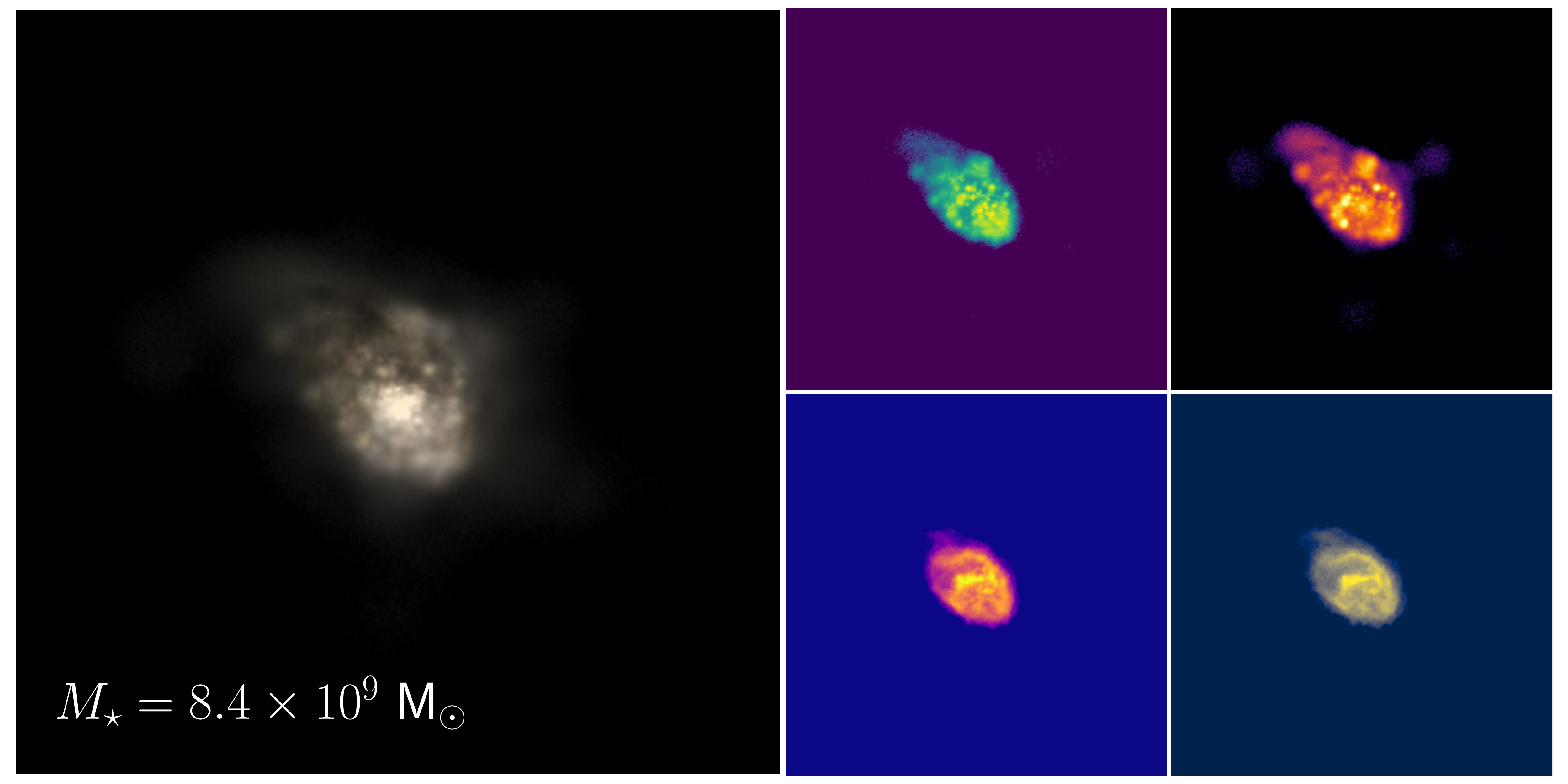}
				\includegraphics[width=0.495\textwidth]{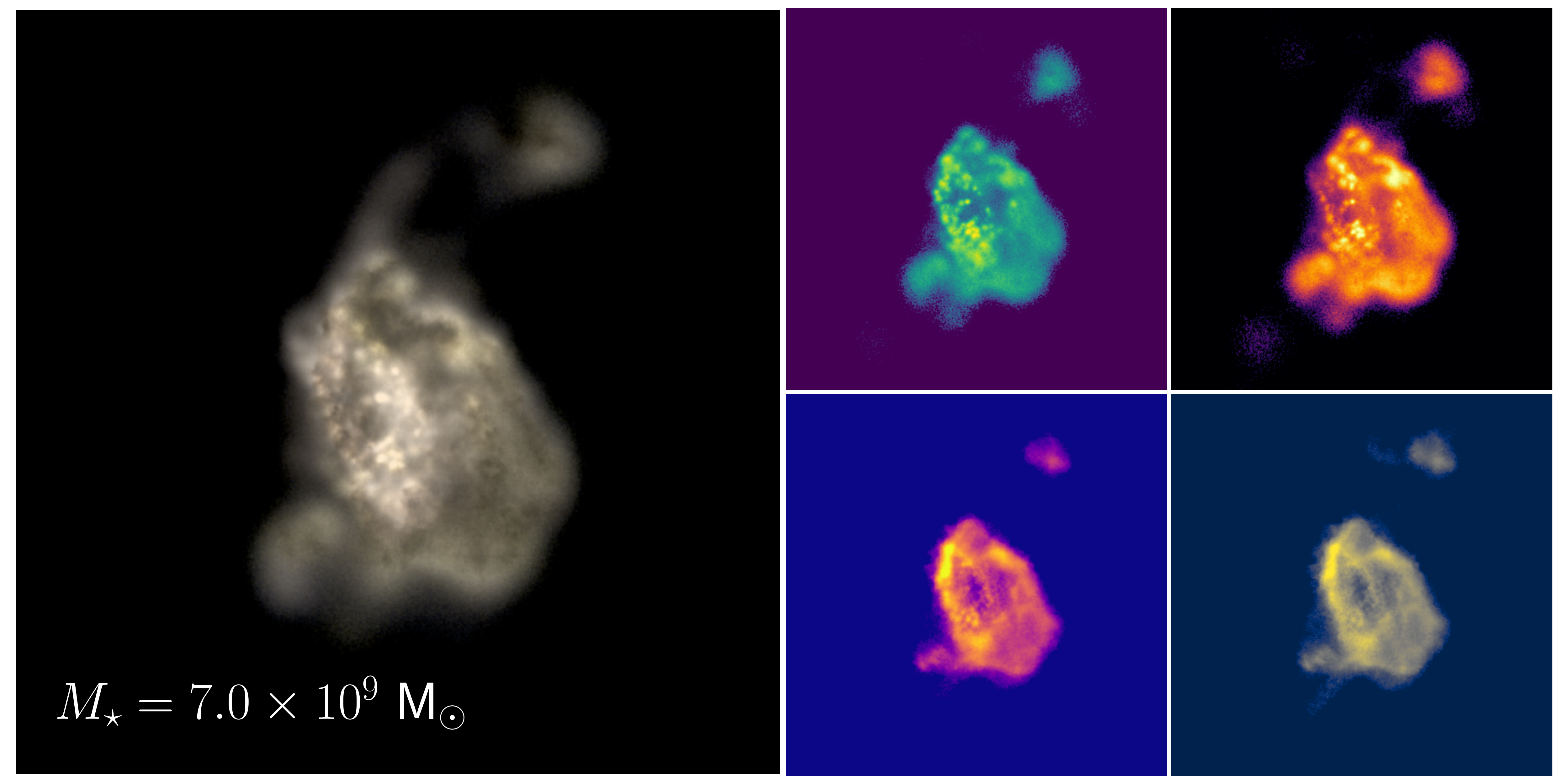}
 \end{center}
    \caption{Face-on images of six massive galaxies in \thesanone at $z=6$. The large panels on the left show mock \texttt{JWST} images constructed for the NIRCam F277W, F356W, and F444W filters. The smaller panels show resolved emission line maps (not continuum subtracted) of \halpha, \oiii, \oire, and \cii. All images cover a $10 \times 10$ kpc$^2$ field of view with $500 \times 500$ pixels. The quoted stellar masses are within two times the stellar half-mass radius.}
    \label{fig:image}
\end{figure*}

Figure~\ref{fig:met} shows the mass-weighted metallicity (in units of the solar metallicity, $Z_\odot$) of all stars less than \mbox{$5$ Myrs} old as a function of the star-formation rate of the galaxy, for resolved, star-forming haloes as defined in the previous paragraph. The metallicities of these stars determine the strength of the metal nebular emission lines as they produce the largest amount of ionizing radiation \citep{Conroy2009, BPASS2017}. The metallicity increases considerably, by almost an order of magnitude, with increasing star-formation rate. The evolution with redshift on the other hand is relatively gentle, with a difference of only about $0.2$ dex at a moderate SFR of $\sim 1\,\text{M}_\odot\,\text{yr}^{-1}$. Finally, Figure~\ref{fig:sed} shows both the intrinsic and dust-corrected SED of the central galaxy in the most massive halo in the \thesanone simulation at $z=6$ ($M_\star(<2r_{\mathrm{half},\star}) \simeq 7.01 \times 10^{9} \mathrm{M}_\odot$). The different coloured vertical bands represent the various target emission lines considered in this work. We have chosen to investigate these particular lines because they are predicted to be some of the most luminous in high-$z$ galaxies and are planned to be observed by current and upcoming LIM experiments. Finally, we note that we have only performed this analysis on the \thesanone simulation and any results presented in this work from other simulations in the \thesan suite are derived on the assumption that the \lsfr scaling relations are converged between the resolutions \citep{KannanThesan}.



\section{Results}
\label{sec:results}

\subsection{Photometric properties}
To illustrate the qualitative features of the radiative transfer calculations, we show in Figure~\ref{fig:image} face-on images of six massive galaxies in \thesanone at $z=6$. The large panels on the left show mock \texttt{JWST} images constructed for the NIRCam F277W, F356W, and F444W filters. The smaller panels show resolved emission line maps (not continuum subtracted) of select lines  \halpha, \oiii, \oire, and \cii as indicated. All images cover a $10 \times 10$ kpc$^2$ field of view with $500 \times 500$ pixels. The visualisations show the structural morphology of these massive galaxies, including the impact of dust and subtle differences in the emergent emission line maps. 

\begin{figure}
	\includegraphics[width=0.99\columnwidth]{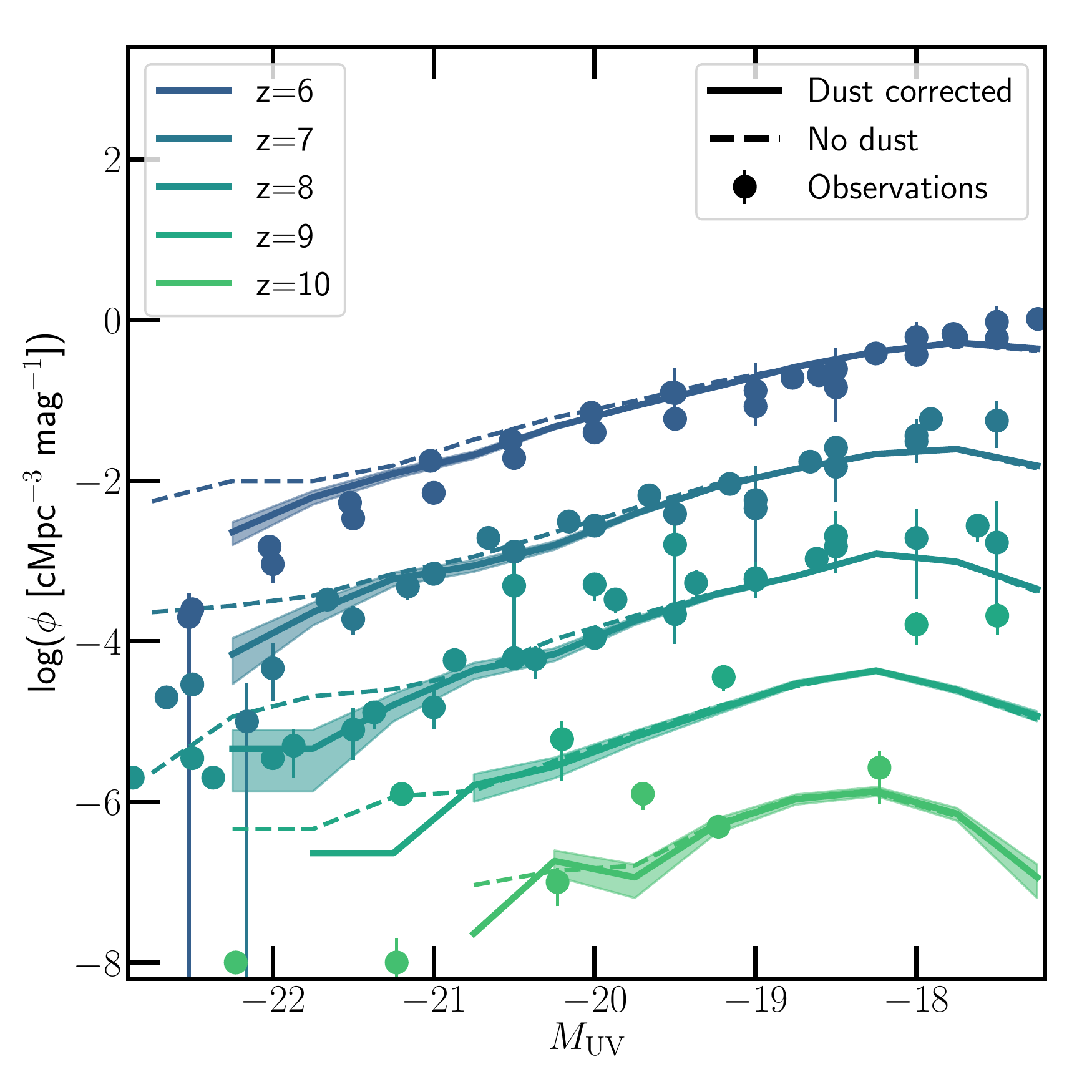}
    \caption{UV luminosity (at rest frame $1500$\,\AA) functions at $z=6$--$10$, for \thesanone derived from the \skirt calculations. The dashed curves show the intrinsic emission of the galaxy while the solid curves are the dust-attenuated values. The observational estimates are taken from \citet{Bouwens2015, Bouwens2017}, \citet{Finkelstein2015}, \citet{McLeod2016}, \citet{Livermore2017}, \citet{Ishigaki2018} and \citet{Atek2018}. We note that the luminosity functions are offset by $\Delta \mathrm{log}(\Phi) = -(z-8)$. The simulated UV luminosity functions match the observational estimates over a wide range of magnitudes.}
    \label{fig:lum}
\end{figure}

A more quantitative analysis of the derived SED and emission line properties of the galaxies in \thesanone begins with Figure~\ref{fig:lum}, which shows the UV luminosity (at rest frame $1500$\,\AA) functions at $z=6$--$10$ as indicated. For comparison, we also show observational estimates (coloured points) from \citet{Bouwens2015, Bouwens2017}, \citet{Finkelstein2015}, \citet{McLeod2016}, \citet{Livermore2017}, \citet{Atek2018} and \citet{Ishigaki2018}. The dashed curves show the intrinsic luminosities of the galaxies, while the solid curves display the dust-attenuated luminosity functions, with the shaded regions indicating the Poisson noise. The simulated luminosity functions start to fall off at \mbox{$M_\mathrm{UV} \gtrsim -18$}, because we chose to model only galaxies above a certain stellar mass and SFR. \citet{KannanThesan} shows that the luminosity functions continue with the same slope up to \mbox{$M_\mathrm{UV} \simeq -12$} in the \thesanone simulation when all the galaxies in the simulation volume are considered. Dust attenuation manages to reduce the UV magnitudes of the most massive star-forming galaxies ($M_\mathrm{UV}\lesssim-21$), such that the simulated estimates match the observational results at almost all redshifts ($z \geq 7$). At $z=6$ however, the model seems to slightly under-predict the dust attenuation at the high mass end, indicating that the DTM ratios might need to be even higher than what is predicted from the scaling relation used in this work. Overall, this figure confirms that our SED modelling give rise to realistic UV luminosities. Moreover, the dust attenuation model does a reasonable job of reducing the UV luminosites of highly star-forming galaxies, such that the simulated UVLF reproduces the observed drop-off at the right magnitudes and by roughly the correct amount.

 \begin{figure}
	\includegraphics[width=0.99\columnwidth]{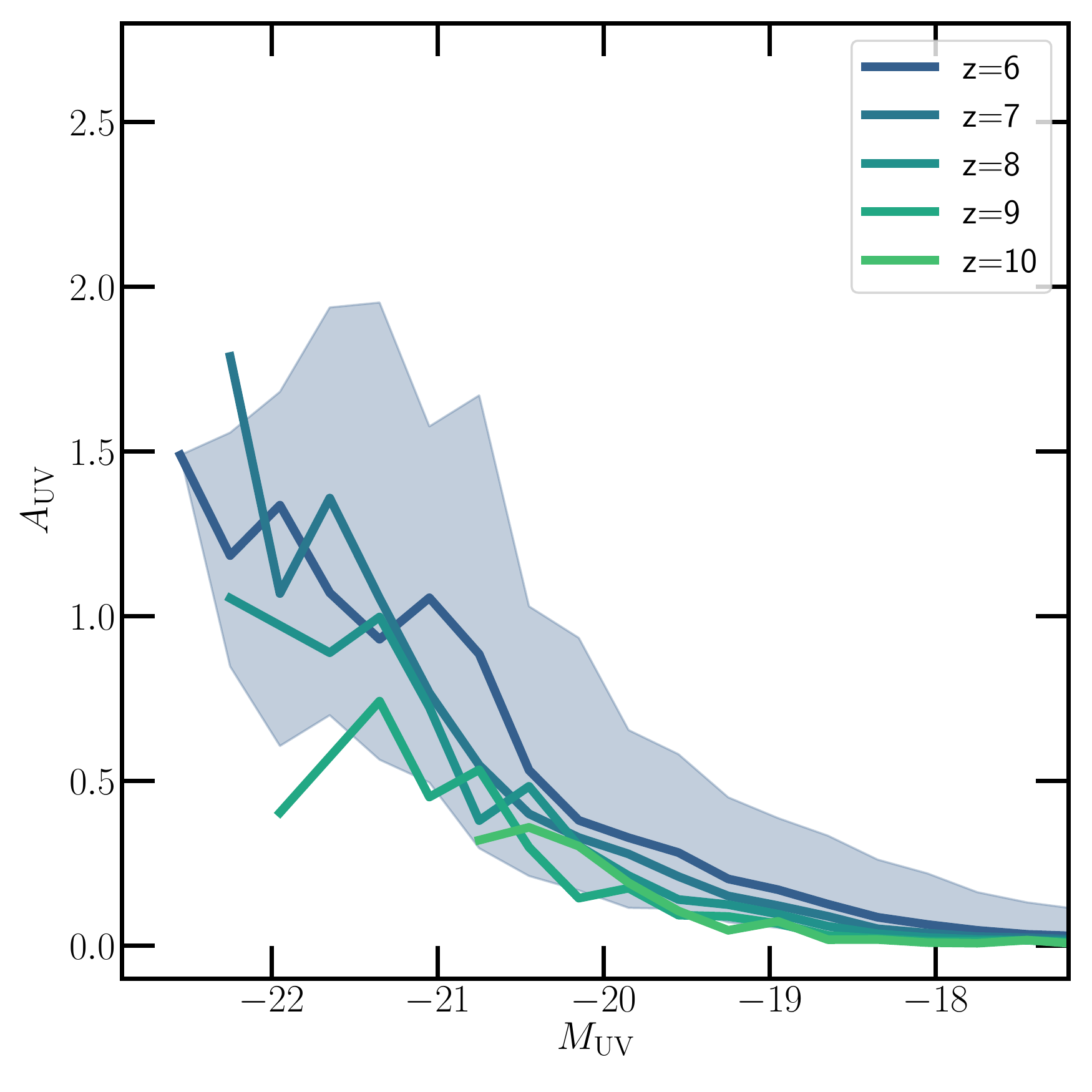}
    \caption{The attenuation, $\mathrm{A_{UV} = -2.5~ log(L_{UV}^{observed}/L_{UV}^{intrinsic})}$, at UV wavelengths as a function of the UV magnitude for the galaxies in \thesanone at $z=6$--$10$ as indicated. The shaded regions shows the $10$--$90$ percentile distribution at $z=6$. While the low luminosity galaxies are largely dust free, the high mass, high SFR, metal enriched galaxies show a steep increase in dust attenuation, especially below $M_\mathrm{UV} \lesssim -21$. This sharp rise in $A_\mathrm{UV}$ is needed in order to match the steep fall off in the observed UVLF happening at about the same magnitude.}
    \label{fig:attn}
\end{figure}

\begin{figure*}
	\includegraphics[width=0.99\textwidth]{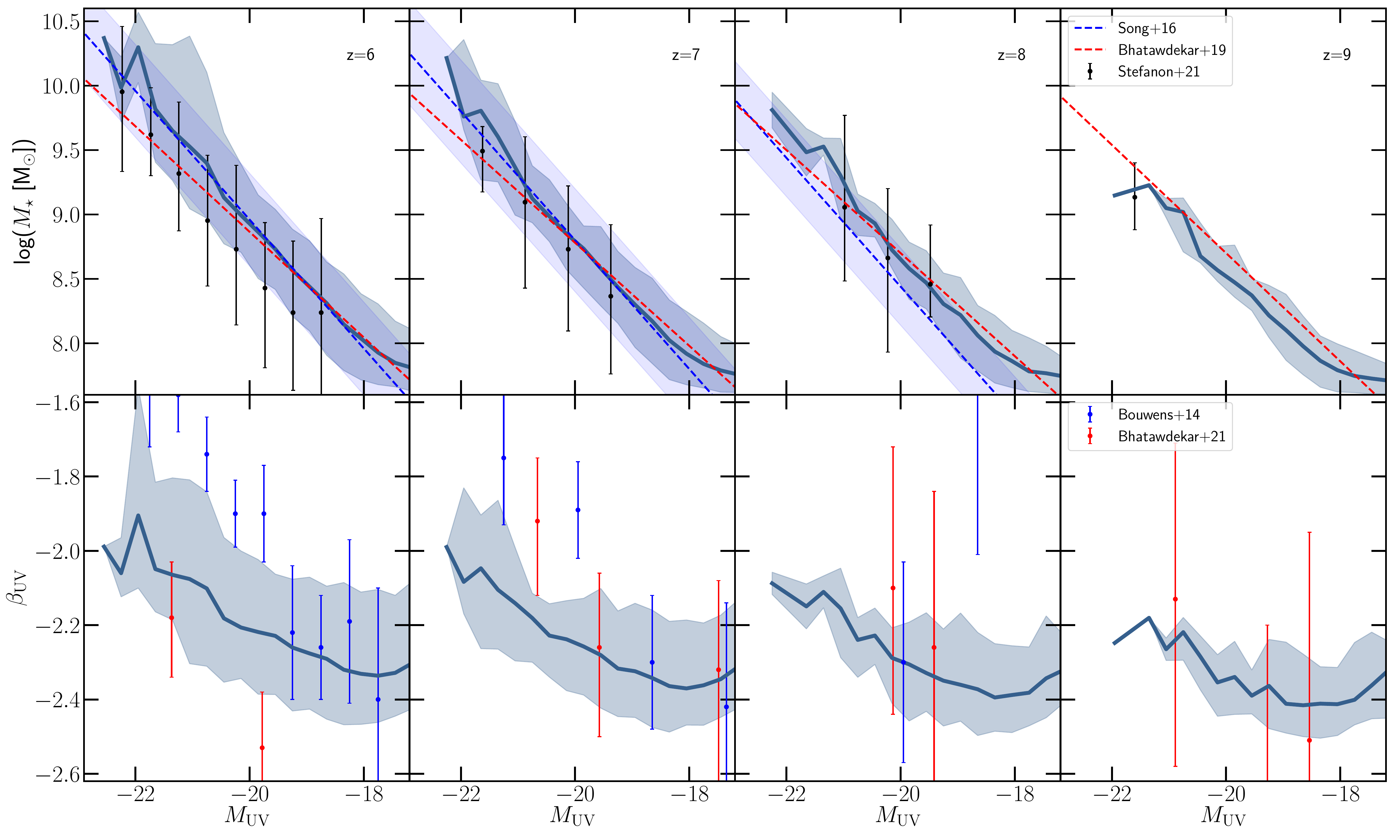}
    \caption{\textit{Top panels}: The stellar mass of the galaxies as a function of the dust-corrected UV magnitude at $z=6$ (first panel), $z=7$ (second panel), $z=8$ (third panel) and $z=9$ (fourth panel). For comparison, we also show observational estimates from \citet[][blue curves]{Song2016}, \citet[][red curves]{Bhatawdekar2019}, and  \citet[][black points]{Stefanon2021}. \textit{Bottom panels}: The UV continuum slope ($1300$--$2650$\,\AA) as a function of the UV magnitude of the galaxies for $z=6$--$9$ as indicated. The shaded regions show the $2\sigma$ scatter in the relation. Observational estimates from \citet{Bouwens2014} and \citet{B21} are shown as blue and red points, respectively.}
    \label{fig:muv}
\end{figure*}

The impact of dust is seen more clearly in Figure~\ref{fig:attn}, which plots the attenuation of the UV flux, $A_\mathrm{UV} = -2.5~\mathrm{log}(L_\mathrm{UV}^\mathrm{observed}/L_\mathrm{UV}^\mathrm{intrinsic})$, as a function of the dust-attenuated UV magnitude of the galaxies at $z=6$--$10$. As expected, the low luminosity ($M_\mathrm{UV} \gtrsim -19$) and by extension low SFR galaxies are least affected by dust. In fact $A_\mathrm{UV} \sim0$, for these low mass haloes, implying that they are effectively dust free. This is because the low SFR haloes are less metal enriched (Figure~\ref{fig:met}), therefore the dust content and corresponding attenuation is lower. As the SFR increases, the galaxies become more metal enriched and dust obscured. Additionally, the slight evolution in the metallicity and DTM ratio with redshift imprints a corresponding evolution in the attenuation factor. At higher redshifts, galaxies have $A_\mathrm{UV}\sim 0.5$ even at $M_\mathrm{UV} \lesssim -21$. On the other hand the attenuation is much more rapid at $z=6$ with an almost exponential increase in the UV attenuation for
galaxies brighter than $M_\mathrm{UV} \lesssim -20$, hinting at a rapid buildup of dusty, high-mass, high-SFR galaxies during the reionization epoch. This steep increase in $A_\mathrm{UV}$ is needed to account for the steep drop-off in the observed UVLF happening at about the same magnitude. The shaded region shows the scatter in this relation, which is quite large especially for the highest luminosity galaxies. The scatter mainly arises because the viewing angle of the galaxy is random. For example, a galaxy viewed edge-on will have significantly more attenuation than one that is face-on. The impact of viewing angle on the attenuation curve of galaxies will be investigated in future works.

In the top panels of Figure~\ref{fig:muv} we show the stellar mass of the galaxies as a function of the dust-corrected UV magnitude at $z=6$--$9$ as indicated. The solid curves show the median and the shaded regions represent the $2\sigma$ scatter around it. The dashed blue curves and black points show the observational data derived using stellar mass estimates from deep IRAC/Spitzer measurements of galaxies in the reionization epoch as outlined in \citet{Song2016} and \citet{Stefanon2021}, respectively. The dashed red lines are the best-fit curves to the stellar masses derived for the galaxies in the Hubble Frontier Fields by combining the \textit{HST} imaging with \textit{Spitzer} and ground-based VLT data \citep{Bhatawdekar2019}. At lower redshifts ($z=6$ and $z=7$), the simulations seems to be in better agreement with the \citet{Song2016} data, while \citet{Bhatawdekar2019} and \citet{Stefanon2021} match the simulated values for $z=8$ and $z=9$ more accurately. The scatter in the relation from the simulations is consistent with the observational scatter.

Along similar lines, the bottom panels of Figure~\ref{fig:muv} present the UV continuum slopes ($\beta_\mathrm{UV}$) as a function of the galaxy UV magnitude at $z=6$ (first panel), $z=7$ (second panel), $z=8$ (third panel) and $z=9$ (fourth panel). The UV continuum depends on the mass and surface temperature of the O \& B stars, therefore, it will change according to the underlying IMF and the metal enrichment within the galaxies. Moreover, the star formation history of a galaxy will determine the number of young stars and, hence, the UV slopes also contain information about the recent star formation episodes within the galaxies. Finally, the UV continuum is also affected by the distribution of dust and their grain size distribution \citep[see for example][]{Reddy2012}. Therefore, this measurement helps constrain a variety of physical properties that govern galaxy formation \citep{Wilkins2013}. The slopes are measured by fitting the dust-attenuated SEDs in the $10$ wavelength filters suggested by \citet{Calzetti1994}, with a power-law function, $f_\lambda \propto \lambda^{\beta_\mathrm{UV}}$. The shaded regions show the $2\sigma$ scatter around the median. The red and blue points show observational estimates from \citet{Bouwens2014} and \citet{B21}, respectively. At $z=6$, the two observational works are in tension with the \citet{Bouwens2014} estimates, showing consistently higher values than the results from \citet{B21}. At higher redshifts, the estimates are more compatible with each other. The simulated low luminosity galaxies ($M_\mathrm{UV} \gtrsim -19$) match the \citet{Bouwens2014} values at $z=6$ and lie above the estimates from \citet{B21}. At higher redshifts they are in general agreement with both the observational estimates. Since most of these galaxies are mainly dust free, these results imply that the underlying IMF and SFRs of the simulated galaxies are in general agreement with the  observationally inferred estimates. The observed UV continuum slopes of massive galaxies diverge, with \citet{Bouwens2014} showing a steep increase in $\beta_\mathrm{UV}$, while \citet{B21} predict a relatively redshift independent evolution. However, we caution that the \citet{B21} data predicts that bright galaxies at $z=6$ are significantly bluer than the bright galaxies at $z=7$, which is contrary to the general picture of galaxies getting more dust enriched with time. Therefore, the bright galaxies in the \thesan simulations seem to show shallow UV continuum slopes that are in tension with the observed data. This might be a consequence of the assumed DTM scaling, which evolves very strongly with redshift, leading to too little dust at these high redshifts. Similar discrepancies have been found in other simulation efforts \citep{Gnedin2014, Shen2020, Wu2020, Vijayan2021} and they may be attributed to disparate extinction curves at high-$z$. An investigation into the variations caused by different extinction curves is beyond the scope of this paper, but we refer the reader to \citet{Shen2020} for a more extensive discussion of this topic. These three plots (Figures~\ref{fig:lum},~\ref{fig:attn} and~\ref{fig:muv}) indicate that the simulations and SED model do a good job of reproducing the observed UV properties and magnitude of dust attenuation in high-redshift galaxies.

\begin{figure*}
	\includegraphics[width=0.99\textwidth]{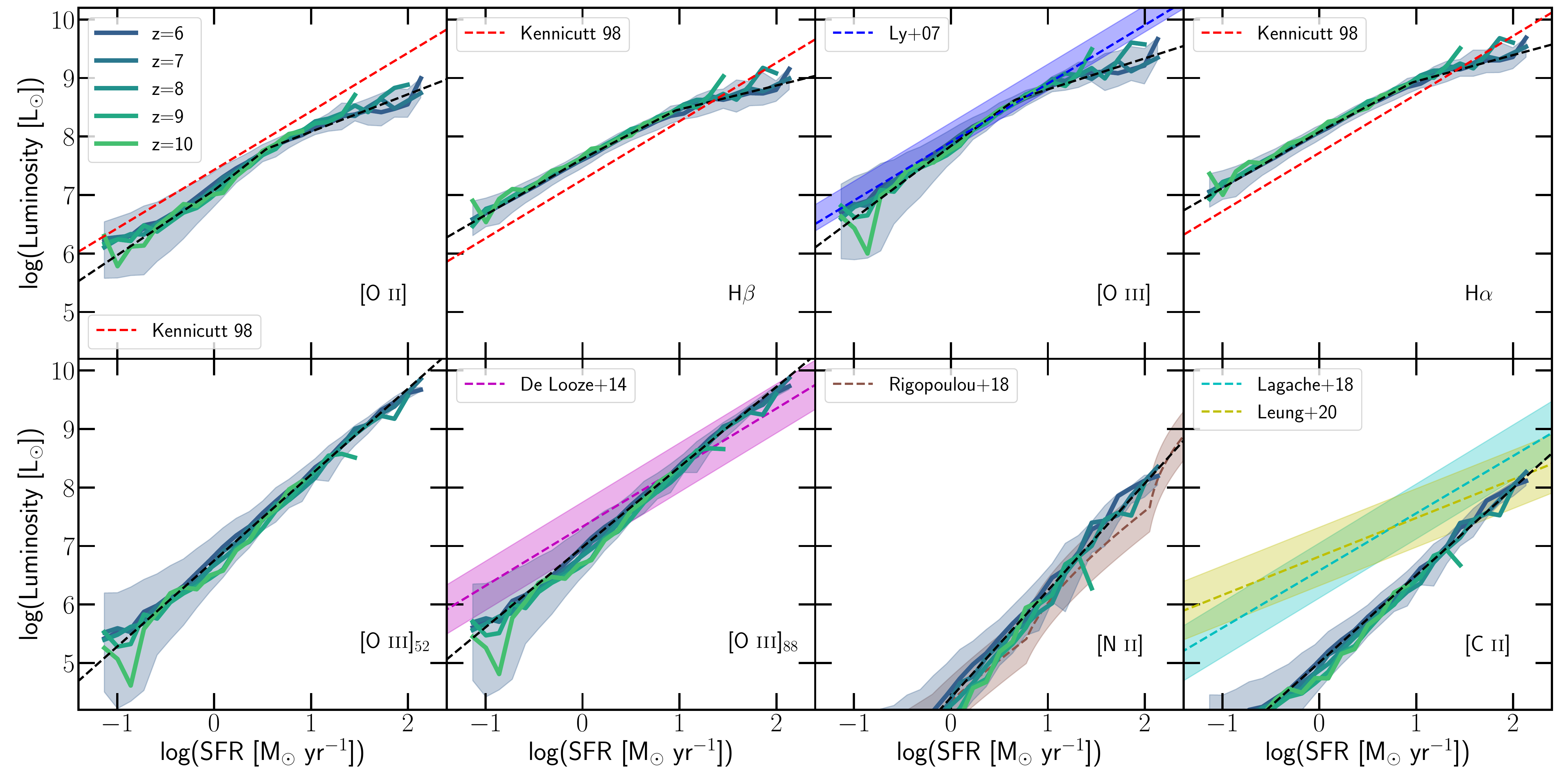}
    \caption{The emission line luminosities of \mbox{\oii (first panel)}, \mbox{\hbeta (second panel)}, \mbox{\oiii (third panel)}, \mbox{\halpha (fourth panel)}, \mbox{\oirf (fifth panel)}, \mbox{\oire (sixth pnael)}, \mbox{\nii (seventh panel)} and \mbox{\cii (eighth panel)} as a function of the SFR of the galaxies at $z=6$--$10$ as indicated. The shaded regions show the $10^\text{th}$ and $90^\text{th}$ percentiles of the distribution at $z=6$. The dashed black curves show the best-fit relation using Eq.~(\ref{eq:fit}). Observational and theoretical estimates from \citet[][dashed red curves]{Kennicutt1998}, \citet[][dashed blue curves]{Ly2007}, \citet[][dashed magenta curves]{DeLooze0214}, \citet[][dashed cyan curves]{Lagache2018}, \citet[][dashed brown curves]{R18} and \citet[][dashed yellow curves]{Leung2020} are also shown for comparison. In general, the predicted emission line luminosities are in good agreement with the estimates found in the literature, although important differences exist.}
    \label{fig:sfrlum}
\end{figure*}

\subsection{Emission line luminosities}

We now shift our focus to measuring the luminosities of the emission lines considered in this work. In Figure~\ref{fig:sfrlum} we plot the line luminosities (in \mbox{L$_\odot$}) of \oii (first panel), \hbeta (second panel), \oiii (third panel), \halpha (fourth panel), \oirf (fifth panel) \oire (sixth panel) \nii (seventh panel) and \cii (eighth panel) as a function of the SFR of the galaxies. The solid lines indicate the median relation from the simulation, with the various colours denoting the different redshifts from $z=6$--$10$ as indicated. The corresponding shaded regions show the $10^\text{th}$ and $90^\text{th}$ percentiles of the distribution at $z=6$. Similar scatter is observed at higher redshifts as well, but this is not shown in order to maintain the clarity of the plot. The median relation shows minimal evolution with redshift for all the lines considered in this work, because of the sluggish evolution in the metallicity of \hii regions with redshift (see Figure~\ref{fig:met}). In fact, the median relations at all redshifts more or less lie on top of each other, with the relative spread much smaller than the scatter in the relations at $z=6$.

We therefore choose to quantify the line luminosities as a function of the SFR alone using a broken power-law fit of the form 
\begin{equation}
    y = \left\{ \begin{array}{ll}
    a + m_a x & x < x_b\\
    a + (m_a - m_b) x_b + m_b x & x_b \leq x < x_c \\
    a + (m_a - m_b) x_b + (m_b - m_c) x_c + m_c x &\text{otherwise}
    \end{array} \right.
    \label{eq:fit}
\end{equation}
where $y=\mathrm{log}(L\,/\,\mathrm{L}_\odot)$ and $x=\mathrm{log(SFR)}$ in units of $\mathrm{M_\odot \, yr^{-1}}$. The fit involves three regimes. First, $x<x_b$, which encapsulates the behaviour at low star-formation rates corresponding to line emission from low metallicity dust-free galaxies. The intermediate regime ($x_b\lesssim x  < x_c$) accounts for the slightly metal and dust enriched galaxies with relatively low dust attenuation. Finally, at high star-formation rates, the galaxies are highly metal enriched, which in turn leads to high dust masses and high levels of attenuation. As the low star-formation rate galaxies (SFR$\lesssim1$~M$_\odot$~yr$^{-1}$) are largely dust free, we set $x_b=0$, while all the other parameters are fit using the method of least squares. Additionally, the far-IR lines are not expected to be attenuated by dust \citep{Draine2007}, so we only use a single power law for these lines. The best-fit values for the different lines and their $1\sigma$ error are summarised in Table~\ref{table:fits}.

\begin{table*}
	\centering
	\caption{Best-fit values and $1\sigma$ error bars for the emission line luminosity as a function of the star-formation rate of the galaxy. The table lists the names (first column) and wavelengths (second column) of the lines considered in this work along with the parameters used to describe the \lsfr relation (columns three to seven) as outlined in Eq.~(\ref{eq:fit}).}
	\label{table:fits}
	\begin{tabular}{llccccc} 
		\hline
		Line & Wavelength (\AA) & $a$ & $m_a$ & $m_b$ & $m_c$ &$x_c$ \\
		\hline
		{[\ion{O}{II}]} & 3726, 3729 & 7.08 $\pm$ 0.006 & 1.11 $\pm$ 0.016 & 1.31 $\pm$ 0.030 & 0.64 $\pm$ 0.031  & 0.54 $\pm$ 0.032 \\ 
 		\ion{H}{$\beta$} & 4861 & 7.62 $\pm$ 0.002 & 0.96 $\pm$ 0.007 & 0.86 $\pm$ 0.008 & 0.41 $\pm$ 0.027  & 0.96 $\pm$ 0.035 \\ 
 		{[\ion{O}{III}]} & 4959, 5007 & 7.84 $\pm$ 0.006 & 1.24 $\pm$ 0.016 & 1.19 $\pm$ 0.027 & 0.53 $\pm$ 0.039  & 0.66 $\pm$ 0.040 \\
 		\ion{H}{$\alpha$} & 6563 & 8.08 $\pm$ 0.002 & 0.96 $\pm$ 0.006 & 0.88 $\pm$ 0.008 & 0.45 $\pm$ 0.025  & 0.96 $\pm$ 0.032 \\
 		{[\ion{O}{III}]}$_{52}$ & $5.2 \times 10^5$ & 6.75 $\pm$ 0.007 & 1.47 $\pm$ 0.013 & -  & - & - \\
		{[\ion{O}{III}]}$_{88}$ & $8.8 \times 10^5$ & 6.98 $\pm$ 0.008 & 1.37 $\pm$ 0.014 & - & - & - \\ 
		{[\ion{N}{II}]} & $1.22 \times 10^6$ & 4.42 $\pm$ 0.010 & 1.83 $\pm$ 0.018 & - & -  & - \\
		{[\ion{C}{II}]} & $1.58 \times 10^6$ & 5.01 $\pm$ 0.005 & 1.49 $\pm$ 0.011 & - & - & - \\ 
		\hline
	\end{tabular}
\end{table*}

The ionized hydrogen gas in \hii regions recombines in a cascading process that produces various strong emission lines including \halpha and \hbeta.  The line luminosities show an almost linear relation with SFR at low SFRs. It then becomes shallower due to increased dust attenuation. Encouragingly, the intrinsic \halpha/\hbeta ratio is about $2.9$ which is very close the theoretical value of $2.87$ that assumes an electron temperature of $T_{e}=10^4~\mathrm{K}$, an electron density of $n_{\rm e}=10^2~\mathrm{cm}^{-3}$, and Case B recombination conditions \citep{Osterbrock2006}. The dashed red curves show the estimates from \citet{Kennicutt1998}, who assumed an optically thick ISM, case B recombination, a \citet{Salpeter1955} IMF and that all UV continuum is absorbed by the gas in the galaxy. The plotted relation has been re-scaled to match a \citet{Kroupa2001} IMF adopted in this work by using the conversion presented in \citet{Driver2013}. The intrinsic luminosities of these Balmer lines (devoid of any dust attenuation) derived from the \thesan simulations is about a factor of $0.3$ dex higher and the slopes are slightly shallower than the values obtained by \citet{Kennicutt1998}. These differences arise due to the fact that the low SFR galaxies also have low stellar metallicities, leading to an increased output of ionizing photons, resulting in a higher \halpha and \hbeta luminosity. In fact, it has been shown that decreasing the metallicity can increase the Balmer line luminosities by about $\sim 0.25$ dex \citep{Leja2017}. This will increase the luminosity of the Balmer emission lines at low SFR while maintaining similar values in galaxies which contain stars that have $Z_\star \sim Z_\odot$, which in our simulations happens only in the rare highly star-forming haloes (Figure~\ref{fig:met}). This effect, therefore, explains the discrepancies seen in both the amplitude and slope of the relation.

Ionized oxygen in the \hii regions produces a variety of emission lines in both the optical and FIR.  We first look at the forbidden \mbox{\oii ($3726, 3729$\,\AA)} doublet. Similar to the Balmer lines, we show the rescaled estimate from \citet[][dashed red curve]{Kennicutt1998}. The simulated values are slightly below the expected relation due to the low metallicity of \hii regions in low SFR galaxies. The luminosities increase with a power law slope greater than one, as the metallicity increases with the SFR. At high SFRs dust attenuation kicks in and reduces the expected luminosities again. Another important oxygen emission line in the visible spectrum is the \mbox{\oiii ($4959, 5007$\,\AA)} doublet. For comparison we show the observational estimates from \citet[][dashed blue curve]{Ly2007}, who derived the relation from $197$ galaxies in the redshift range $0.07$--$1.47$ from the Subaru Deep Field. Akin to the \oii scaling relation, the line luminosities at low SFR are lower than the expected values due to the low metallicity of \hii regions and then increase with a super-linear power law slope until dust attenuation becomes significant and reduces the luminosities again. 

The far-infrared (FIR) fine-structure lines of carbon, nitrogen and oxygen are among the most dominant coolants in the ISM of galaxies. They are excellent extinction-free probes of the physical conditions of the gas and the intensity and hardness of the ambient interstellar radiation fields \citep{FO388}. We start with the  \oiii lines at $52$\,\micron \xspace and $88$\,\micron \xspace (bottom left panels of Fig.~\ref{fig:sfrlum}). Since the ionization potential for doubly ionized oxygen is $35$ eV, the \ion{O}{III} regions reside close to the hot and young O stars. This makes the \oiii emission co-spatial with the other strong nebular emission lines like the Balmer lines making them important tracers of the star formation in the galaxy. Overplotted is the observed relation for \oire from \citet{DeLooze0214} who derived the relation using a sample of low-metallicity dwarf galaxies from the \textit{Herschel} Dwarf Galaxy Survey and a broad sample of galaxies of various types and metallicities from the literature. The simulated relation appears to be slightly steeper than the observed one, however, they do match over a wide range in SFR ($0<\mathrm{SFR[M_\odot\,yr^{-1}]}<2$) within the errorbars, but disagree at very low metallicities/SFRs by about $0.3$ dex. The luminosity of the \oirf line is similar to that of \oire, mainly because the line ratio is about $\sim 1$ in most environments and is completely insensitive  to the metallicity of the \hii region and is only slightly dependent on the density of the ionized cloud \citep{Yang2021}.

Another important nebular emission line is the fine structure line of the singly ionized nitrogen atom, \nii at rest frame $122$\,\micron. Photons of energy greater than $14.5$ eV are needed to ionize the neutral N atom, implying that the \nii line emission mainly arises in highly photoionized \hii regions around old O and young B stars \citep{FN2}.  The simulated median  relation (third panel from the left in the bottom row of Figure~\ref{fig:sfrlum}) shows a very steep power law slope of around $\sim 1.8$, which is in disagreement with observations from \citet{Spinoglio2012} who estimate a power law slope of $\sim 1$. However, these observations were made at low redshifts and do not fully account for the low metallicities of high-redshift galaxies. It has recently been shown that the \nii luminosity falls off rapidly with decreasing metallicity and can therefore be used as a metallicity indicator for high-redshift galaxies \citep{R18}. We therefore estimate the expected $L_{[\ion{N}{II}]}$--SFR relation by combining the $L_{[\ion{O}{III}]_{88}}$--SFR relation of \citet{DeLooze0214} with the metallicity dependant \oire / \nii line ratio derived in \citet{R18}. We specifically consider models with the ionization parameter $U_\mathrm{ion} = 10^{-2}$, which matches the parameters used in this work. The metallicity at a particular SFR is taken self-consistently from the simulations as shown in Figure~\ref{fig:met}. The resulting relation is shown as a brown curve in the panel. The simulated values and the derived relation lie on top of each other, thereby highlighting the importance of understanding line emission luminosities in low metallicity environments in order to make accurate predictions for line intensity mapping studies that target the Epoch of Reionization. 

Finally, we turn our attention to luminosities for the $158$\,\micron \xspace line emitted by the singly ionized C$^+$ atom. The relatively low ionization potential of carbon ($\sim 11.2$\,eV) makes C$^+$ one of the most abundant metal ions in a variety of environments. Its fine structure line, \cii,  provides the most efficient cooling mechanism for gas in PDRs. This line is also emitted from ionized regions, cold atomic gas and CO dark clouds \citep{Olsen2015}. It is therefore often the strongest IR emission line in galaxy spectra. Moreover, this line is easily accessible from the ground at $4.5\lesssim z \lesssim8.5$. The extensive abundance of \cii emission line environments makes predicting the line luminosities extremely difficult. An accurate estimate will require robust \hii region modelling coupled with self-consistent temperature and radiation field intensities in the PDRs. These in turn will depend on the metal and dust distributions within galaxies. Therefore fully-coupled RHD simulations, that resolve the multi-phase structure of the ISM \citep{Pallottini2019, Katz2019, Lupi2020} are required to properly model \cii line luminosities. However, these simulations are computationally expensive restricting our ability to make simultaneous predictions in a variety of environments.

 \begin{figure*}
	\includegraphics[width=0.99\textwidth]{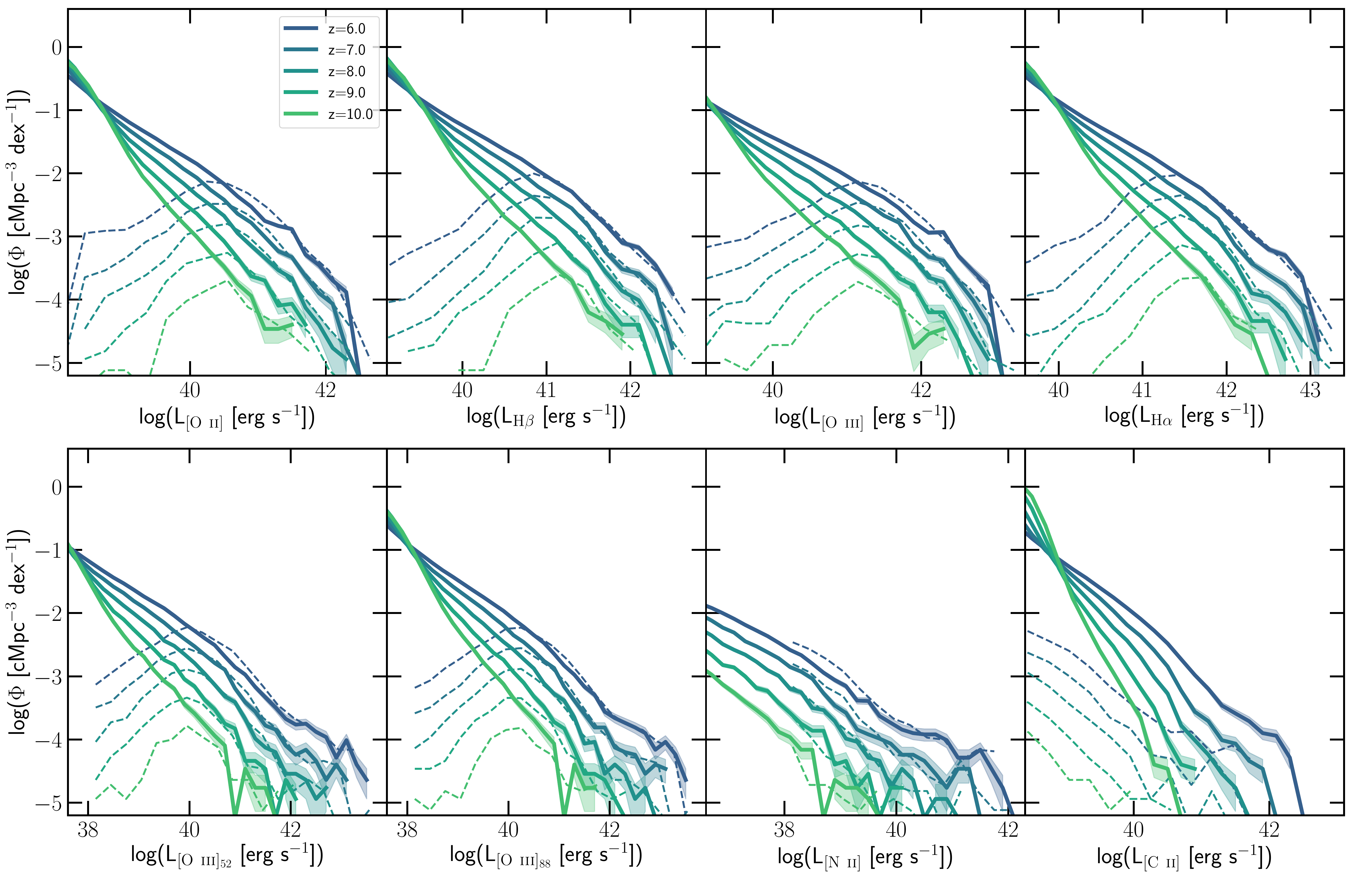}
    \caption{The line luminosity functions of the eight emission lines considered in this work at $z=6-10$ as indicated. The dashed curves show the luminosity functions derived using the values obtained by the SED modelling  technique described in Section~\ref{sec:methods}. The decline at low luminosities arises from the selection function that only includes emission from haloes that are `resolved' \mbox{($M_\star \gtrsim 3 \times 10^7$ M$_\odot$)} and have relatively high star-formation rates \mbox{($\gtrsim 0.1$ M$_\odot$ yr$^{-1}$)}.  The solid curves, on the other hand, are luminosity functions estimated from the best-fit relations (outlined in Table~\ref{table:fits}) applied to all identified subhaloes in the simulation with a non-zero star-formation rate. }
    \label{fig:lumfunc}
\end{figure*}

Recent theoretical efforts have focused on using sub-resolution \hii region and PDR modelling of the galaxies in semi-analytic models \citep[][dashed cyan curve]{Lagache2018} and hydrodynamic simulations \citep[][dashed yellow curves]{Leung2020} to predict \cii emission line luminosities in a wide variety of environments. Our estimates fall below the luminosity predictions from earlier works, especially in low SFR galaxies, most likely because we are unable to properly model the PDR and CO dark molecular regions. This will require a realistic model for the far-UV radiation field intensity in the galaxy and correct gas temperatures in the PDR regions, which depends on the photoelectric heating rate, which in turn will depend on the dust content. We plan to improve our estimates in a future work using RHD simulations that self-consistently model the dust content, molecular gas and radiation field intensities in galaxies \citep{Kannan2020b, Kannan2021}. In particular, this will be part of the \textsc{thesan-zooms} simulation suite that simulates galaxies with a wide range of halo masses ($10^8$--$10^{12} \, \mathrm{M}_\odot$). For the current work, however, we assume that the \cii--SFR relation given by \citet{Lagache2018} is sufficiently accurate and proceed with further analysis.

Finally, in Figure~\ref{fig:lumfunc} we show the line luminosity functions of the eight emission lines considered in this work at $z=6$--$10$ as indicated. For comparison, the dashed curves show the luminosity functions derived using the values obtained by the SED modelling technique described in Section~\ref{sec:methods}. The decline at low luminosities arises from the incomplete selection function that only models emission from haloes that are `well-resolved' \mbox{($M_\star \gtrsim 3 \times 10^7$ M$_\odot$)} and have relatively high star-formation rates \mbox{($\gtrsim 0.1$ M$_\odot$ yr$^{-1}$)}.  The solid curves, on the other hand, are luminosity functions estimated from the best-fit relations (outlined in Table~\ref{table:fits}) applied to all identified subhaloes in the simulation with a positive star-formation rate. The star-formation rate of a galaxy is calculated by summing up all the SF probabilities of the cells in the equation of state \citep[see][for more details]{Springel2003}. This allows us to get an accurate estimate of the current star-formation rate of a galaxy and gets rid of the inherent noise arising from the probabilistic nature of the star formation and star particle creation routine. The fitting functions do a good job of reproducing the luminosity function at the high luminosity end where the selection function is complete. More importantly they are able to make predictions for luminosities of the various emission lines down to the resolution limit of the simulation. This is especially important for making LIM predictions as they are generally sensitive to all sources of emission, including low-mass galaxies. We note that the extrapolated luminosity functions of \cii do not match the predicted ones from SED modelling because we choose to show luminosity functions using the relation from \citet{Lagache2018}, instead of the derived fits from our model, which we know to be inaccurate. 

\subsection{Line intensity maps and auto-correlation spectra}

 \begin{figure*}
	\includegraphics[width=0.99\textwidth]{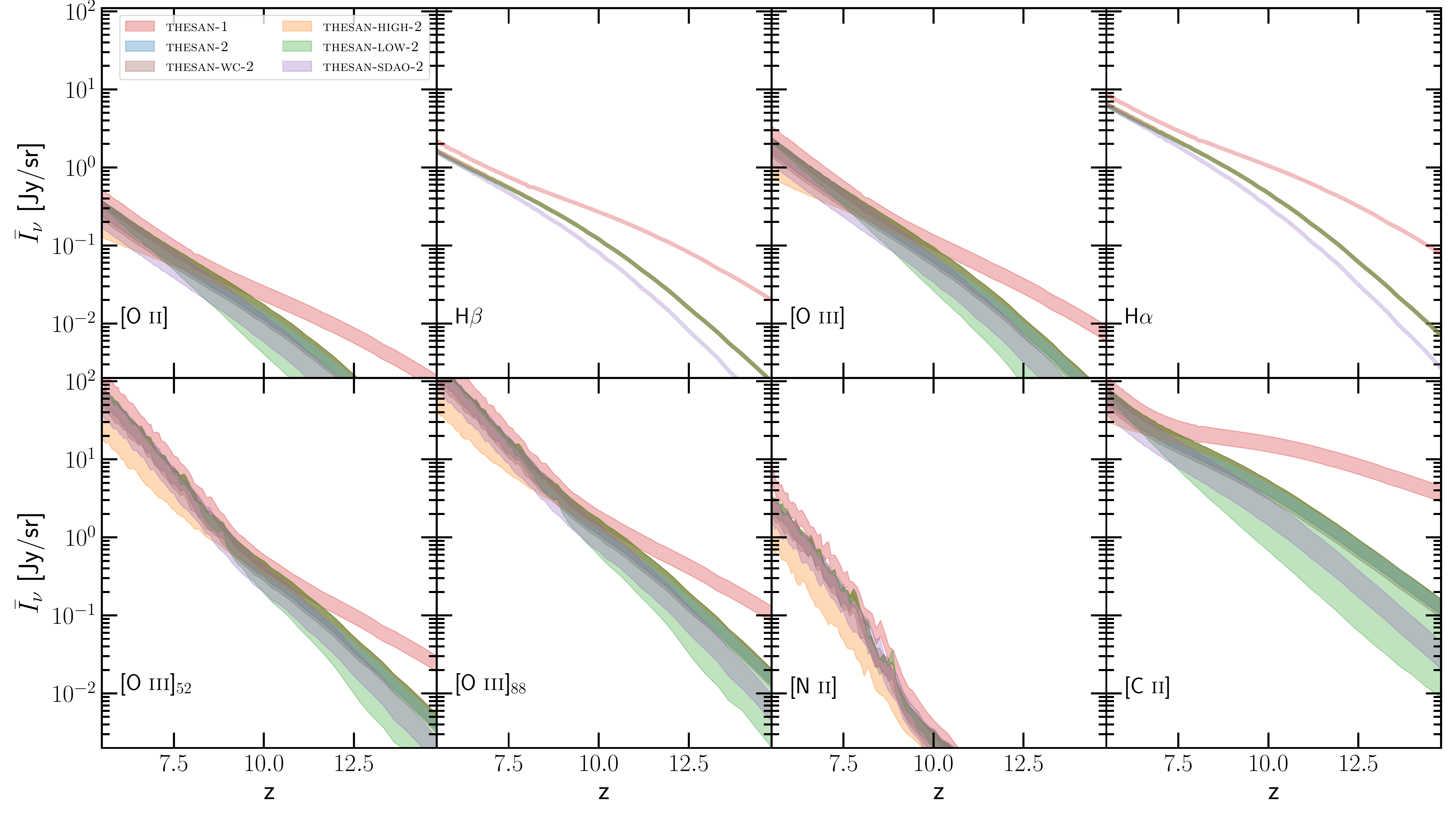}
    \caption{Mean intensity (defined in Eq.~\ref{eq:MI}) of the eight emission lines considered in this work, as a function of redshift. The different coloured lines shows the estimates for \thesanone (red curves), \thesantwo (blue curves), \thesanwc (brown curves), \thesanhigh (orange curves), \thesanlow (green curves) and \thesansdao (purple curves). The shaded regions indicate the uncertainty in the intensity arising from the unknown fraction of LyC photons escaping from nascent \hii regions. The mean intensity of all the lines increases towards lower redshifts mirroring the increase in the global star-formation rate density.}
    \label{fig:MI}
\end{figure*}

The fitting functions derived in the previous section allow us to generate emission line maps based solely upon the position, peculiar velocity \citep[to account for redshift space distortions;][]{Mao2012,Qin2022} and star-formation rate of the galaxy in the simulated volume.\footnote{See Appendix~\ref{app:extrp} for a discussion on the validity of this statement.} We start with Figure~\ref{fig:MI}, which shows the average observed signal or mean intensity ($\bar{I}_\nu$) of the various emission lines as a function of redshift. Specifically, $\bar{I}_\nu$ is defined as \citep{Fonseca2017}
\begin{equation}
    \bar{I}_\nu = \frac{c}{4 \pi} \frac{1}{\nu H(z)} \frac{\sum L_\nu(\mathrm{SFR})}{V_\mathrm{box}} \, ,
    \label{eq:MI}
\end{equation}
where $c$ is the speed of light, $\nu$ is the frequency of the emission line in the rest frame of the galaxy, $H(z)$ is the Hubble constant at redshift $z$, $L_\nu$ is the luminosity of the emission line and $V_\mathrm{box}$ is the comoving volume of the simulation box. The different colours correspond to the different simulations in the \thesan suite, with the red, blue, brown, orange, green and purple curves showing results from \thesanone, \thesantwo, \thesanwc, \thesanhigh, \thesanlow and \thesansdao simulations, respectively. 

The shaded regions represent the uncertainty in the  mean intensity arising from the unknown fraction of LyC photons that escape the \hii regions. While, the emission line analysis assumes that the escape fraction is zero, we know that this cannot be true because such an assumption would not reionize the Universe. In the \thesan simulations, an escape fraction ($f_\mathrm{esc}$) is applied to LyC photons to mimic these processes occurring below the grid scale of the simulation. Multiplying the emission line luminosity of the galaxy with $1-f_\mathrm{esc}$ gives us a lower limit on the amount of emission from these galaxies, while assuming all the LyC emission is absorbed and converted to emission lines from ionized gas sets an upper limit. The mean intensity of the hydrogen emission lines is unaffected by the escape of LyC photons because these photons will eventually interact with a neutral hydrogen atom in the Universe, ionizing it\footnote{We note that this is not necessarily true in the post-reionization Universe, when the mean free path is very large and the time between photon emission and interaction can be substantial. However, based on Fig.~3 of \citet{SmithThesan} there is no indication of significantly deviating from the global budget expected by assuming $f_\text{esc} \approx 0$.}. So the total mean intensity will remain the same but it will alter the spatial distribution of the emission lines, which in turn will imprint a change in the power spectrum.

  \begin{figure*}
	\includegraphics[width=1.03\textwidth]{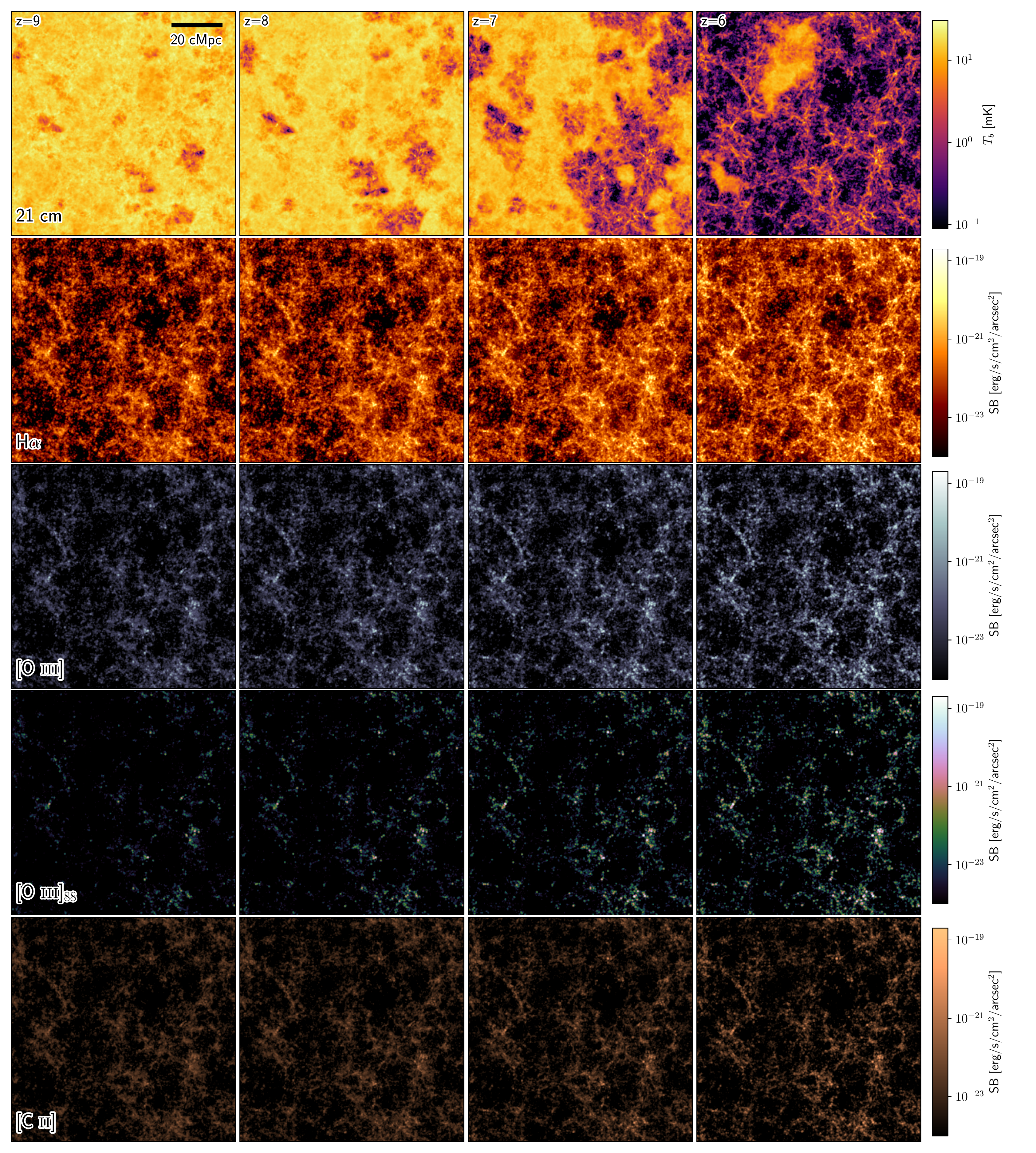}
    \caption{The 21\,cm emission maps (first row) in \thesanone at $z=9$ (first column), $z=8$ (second column), $z=7$ (third column) and $z=6$ (fourth column) compared to the surface brightness maps of the strong nebular emission lines like \halpha (second row), \oiii (third row), \oiii$_{88}$ (fourth row) and \cii (fifth row). The 21\,cm emission starts off quite uniform and gradually diminishes in the ionized regions as reionization progresses. Most of the nebular emission lines are concentrated in the high density filaments which host the star-forming galaxies. \halpha has an additional diffuse component that arises from the radiative recombination of ionized hydrogen throughout the Universe.}
    \label{fig:LIM}
\end{figure*}

 \begin{figure*}
	\includegraphics[width=0.99\textwidth]{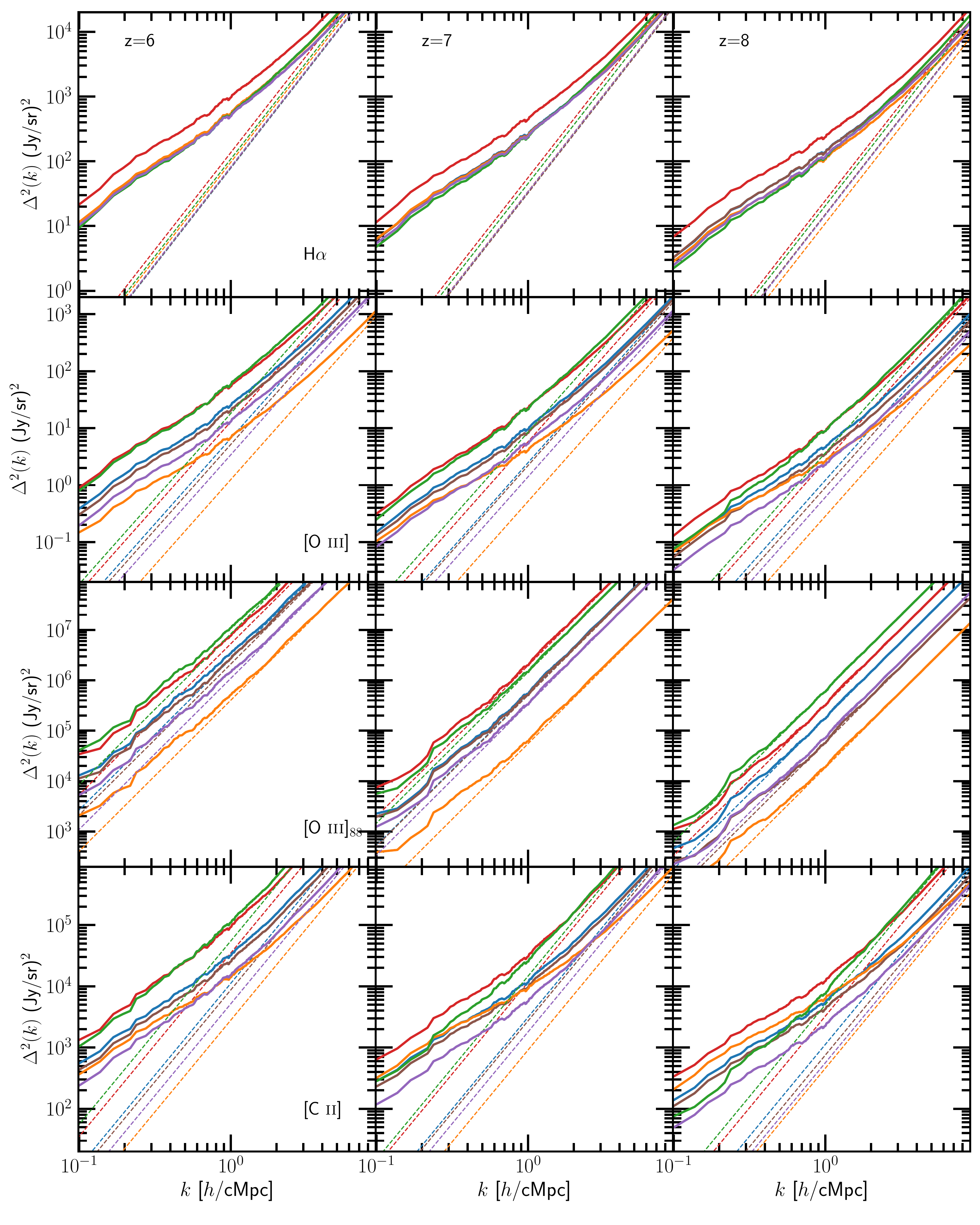}
    \caption{The power spectra of the \halpha (first row), \oiii (second row), \oiii$_{88}$ (third row) and \cii (fourth row) emission lines at $z=6-8$, as indicated.  The solid lines show the power spectrum while the dashed curves indicate the shotnoise contribution to it. The different coloured lines indicate the different simulations in our suite, namely, \thesanone (red curves), \thesantwo (blue curves), \thesanwc (brown curves), \thesanhigh (orange curves), \thesanlow (green curves) and \thesansdao (purple curves).}
    \label{fig:PS_NEL}
\end{figure*}

We note that, while the \lsfr relation has been derived from emission line modelling of \thesanone galaxies, we make the assumption that the relation holds true for the other simulations in our suite as well and calculate the intensity based on just the SFR of the galaxies. This assumption is valid because the resolved galaxy properties are relatively well converged with respect to resolution \citep{KannanThesan}. The mean intensity rises with decreasing redshift  mirroring the increase in the global star-formation rate density \citep{Madau2014}. The predicted intensity of the lines at $z \lesssim 8$ is similar in all the simulations modulo the differences in the escape fractions. At higher redshifts, however, \thesanone predicts significantly stronger emission line intensities because of the increased star-formation rate in low mass haloes. The mean intensities of the metal emission lines show a steeper evolution with redshift than the Balmer lines, because of the additional dependence on the metallicity that decreases with the SFR of the galaxy. We note that the estimated mean intensities of the \oii and \oiii lines are roughly consistent (within a factor of $2$) with the \citet{Gong2017} estimate if extrapolated beyond $z\sim5$. The Balmer emission lines, however, are a factor of $\sim3-4$ higher, due to the increased emission from low metallicity galaxies and the slightly higher star formation rate density. It is quite clear that even measurements of this zeroth order quantity in conjunction with the global star-formation rate density will provide important insights into the \lsfr relation at these redshifts which will in turn place important constraints on the metal content and LyC production rate of these early galaxies.

In Figure~\ref{fig:LIM} we show the spatial distribution of the 21\,cm emission (first row) compared to the surface brightness maps of the four brightest lines considered in this work, namely, \halpha (second row), \oiii (third row), \oiii$_{88}$ (fourth row) and \cii (fifth row) at $z=6$--$9$ as indicated in the \thesanone simulation. The 21\,cm emission is fairly uniform in the beginning, because almost all the gas in the Universe is neutral. At these redshifts the fluctuations in the 21\,cm emission arise from the underlying fluctuations in the matter density field. As reionization progresses, the fluctuations increase as the ionized regions become dark. Eventually, all the gas in the Universe stops emitting in 21\,cm, except for the gas in the high density filaments and galaxies that are self-shielded against the background radiation field. The substructures are quite important because they contribute almost $50\%$ to the 21\,cm power spectrum at scales $k \sim 0.1$--$1$~$h$/cMpc \citep{Kaurov2016}. These scales are expected to be the most foreground-free in observations \citep{HERA}. It is therefore important for reionization simulations to resolve these features in order to accurately predict the 21\,cm power spectrum (PS), especially during the end stages of the reionization process. Simulations that achieve a very low resolution for the radiative transfer calculations in the IGM ($\sim 200$--$330$~ckpc; \citealt{Iliev2014, Hassan2021, Heneka2021}) will not be able to properly resolve these filamentary structures, and will therefore yield inaccurate auto- and cross-correlation power spectra of the 21\,cm line.

In the second row of panels in Figure~\ref{fig:LIM} we show the \halpha emission, both from galaxies and diffuse emission from the recombination of the ionized hydrogen in the IGM. The \halpha radiative recombination luminosity in these resolved emission regions is given by
\begin{equation}
    L_{\halpha}^\mathrm{rec} = h\nu_\halpha \int P_\mathrm{B}\,\alpha_\mathrm{B}(T)\,n_e\,n_p~\mathrm{d}V~,
\end{equation}
where $h\nu_\halpha=1.89$~eV, $P_\mathrm{B}$ is the conversion probability per recombination event, which for \halpha is approximately $0.45$ \citep{Storey1995}, $\alpha_\mathrm{B}$ is the case B recombination rate, $n_e$ and $n_p$ are the number densities of electrons and protons respectively, and $\mathrm{d}V$ is the volume of the cell. We calculate this quantity only for the diffuse gas in the simulation which is not on the equation of state \citep{Springel2003}. We note that the diffuse emission contributes only about $10$--$15\%$ of the total emission in the simulated volume. The major contribution is from ionized \hii regions around newly formed stars in galaxies. The maps clearly show that the most luminous regions of \halpha emission are centred around high density filaments and knots that host galaxies, but there is clearly a non-negligible diffuse emission component that is co-spatial with the ionized regions that are dark in 21\,cm emission. The third, fourth, and firth rows show the emission line maps of \oiii, \oiii$_{88}$, and \cii, respectively. We assume that all of the emission arises from \hii regions in galaxies and that the metal content outside galaxies is low enough that the diffuse component of this emission can be neglected. Additionally, we assume that only $f_\mathrm{abs}=1-f_\mathrm{esc}$ ($0.63$ for \thesanone) of the emitted LyC radiation is able to interact with its immediate surroundings, producing \hii regions while the rest escapes and does not lead to any further ionization of metals in any significant manner. This implies that the emission luminosity of these metal lines is decreased by a factor of $f_\mathrm{abs}$. The lack of a diffuse component concentrates all the emission into the high density filaments.

A more quantitative description of the emission line maps is shown in Figure~\ref{fig:PS_NEL}, which plots the power spectra of the \halpha (first row), \oiii (second row), \oiii$_{88}$ (third row) and \cii (fourth row) emission lines at $z=6$--$8$, as indicated. The solid lines show the power spectrum while the dashed curves indicate the shotnoise contribution to it. The different coloured lines indicate the different simulations in our suite, namely, \thesanone (red curves), \thesantwo (blue curves), \thesanwc (brown curves), \thesanhigh (orange curves), \thesanlow (green curves) and \thesansdao (purple curves). The shotnoise dominates the total power above \mbox{$k \gtrsim 2\,h$/cMpc} for all the lines except \oiii$_{88}$ which is shotnoise dominated all the way down to \mbox{$k \simeq 0.5\,h$/cMpc}. This is because, the steeper \lsfr slope concentrates a large fraction of the total emission in the highly star forming haloes, creating an exceedingly clustered spatial distribution of \oiii$_{88}$, which in turn introduces more discretisation noise on relatively small spatial scales. We note that $f_\mathrm{abs}=1$ for the high (low) mass haloes in the \thesanlow (\thesanhigh) simulation by construction. The difference in the normalization of the PS between different reionization models mainly arisies from the distinct `$f_\mathrm{abs}$' used to match the evolution of the neutral hydrogen density in the Universe. There are some noticeable differences in the shape of the PS of metal emission lines, especially between the \thesanhigh and \thesanlow simulations, arising from the contrasting underlying source populations in these simulations. It is therefore probable that future measurements of the PS could inform us about the escape fraction of LyC photons in addition to constraining the star-formation rates in these early galaxies. 

 \begin{figure}
	\includegraphics[width=0.99\columnwidth]{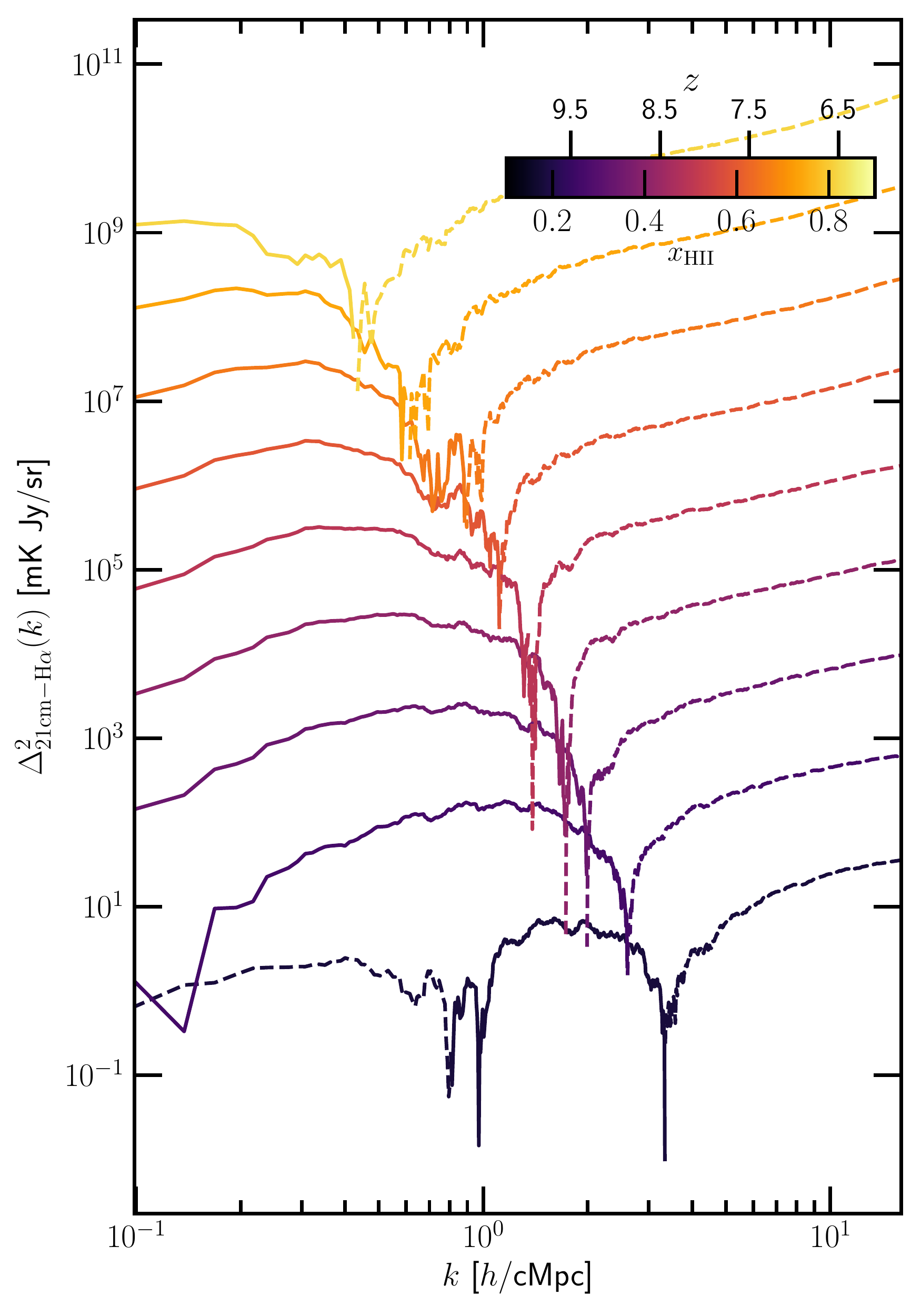}
    \caption{Cross-correlation power spectrum between 21\,cm and \halpha emission in \thesanone at different volume weighted ionized fractions ranging from $0.1$--$0.9$, in increments of $0.1$. The PS are shifted by a factor of $10.0^{(x_\ion{H}{II}/0.1 - 1)}$ for clarity. The dashed curves exhibit a positive correlation while the solid lines indicate negative correlation. This perspective clearly shows that the transition wavenumber ($k_\mathrm{transition}$; defined as the scale at which the correlation function changes sign), moves to larger spatial scales as reionization progresses.}
    \label{fig:cross_Ha_evolve}
\end{figure}

 \begin{figure*}
	\includegraphics[width=0.995\textwidth]{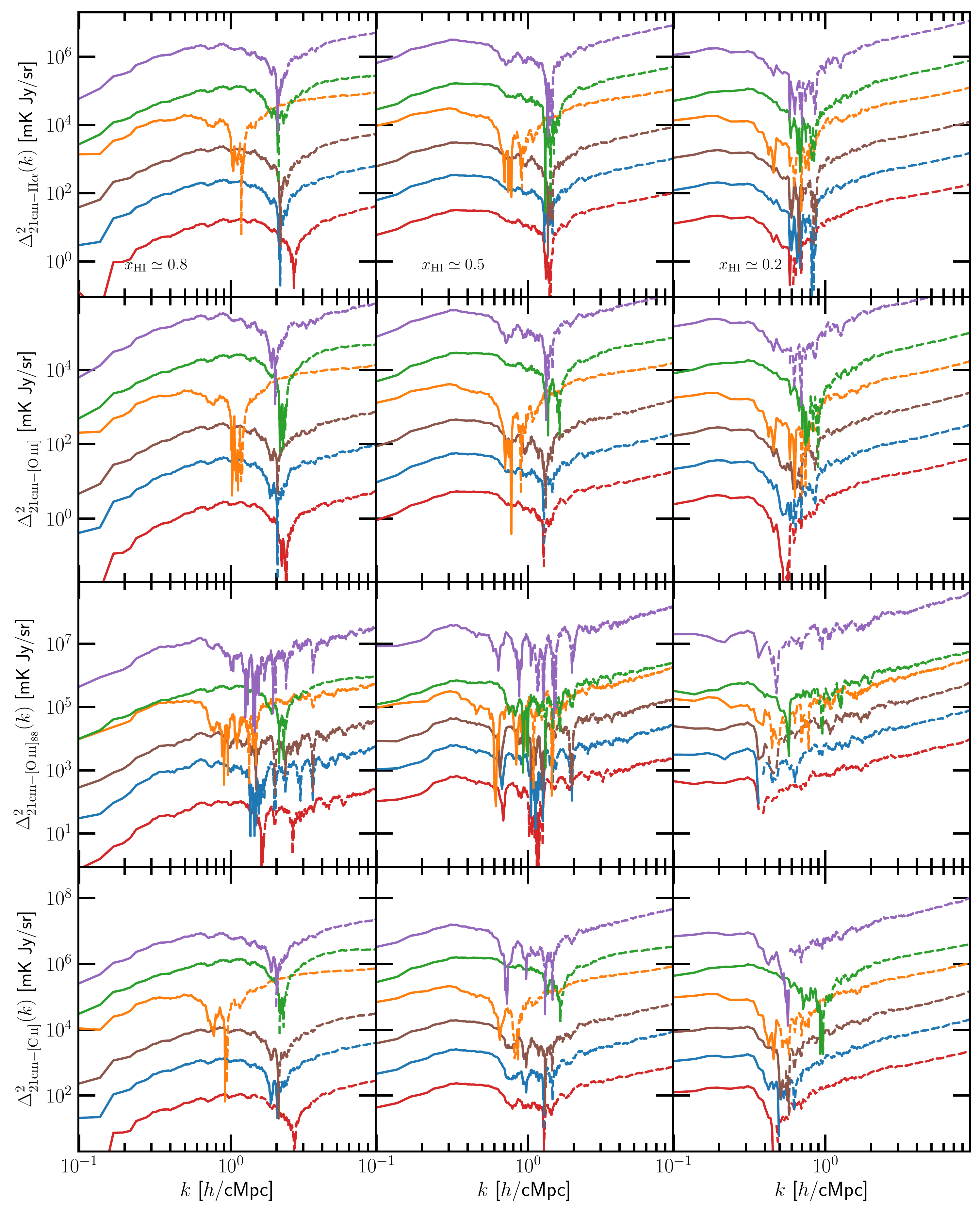}
    \caption{Cross-correlation functions between 21\,cm and \halpha (top row), \oiii (second row), \oiii$_{88}$ (third row) and \cii (fourth row) at volume weighted neutral fractions of $0.8$ (left panels), $0.5$ (middle panels) and $0.2$ (right panels). The different coloured lines indicate the different simulations in our suite, namely, \thesanone (red curves), \thesantwo (blue curves), \thesanwc (brown curves), \thesanhigh (orange curves), \thesanlow (green curves) and \thesansdao (purple curves). The lines are offset by regular intervals (a factor of 10 between each simulation) in order to improve clarity by preventing the lines from overlapping with each other.}
    \label{fig:crossAll}
\end{figure*}

 \begin{figure*}
	\includegraphics[width=0.999\textwidth]{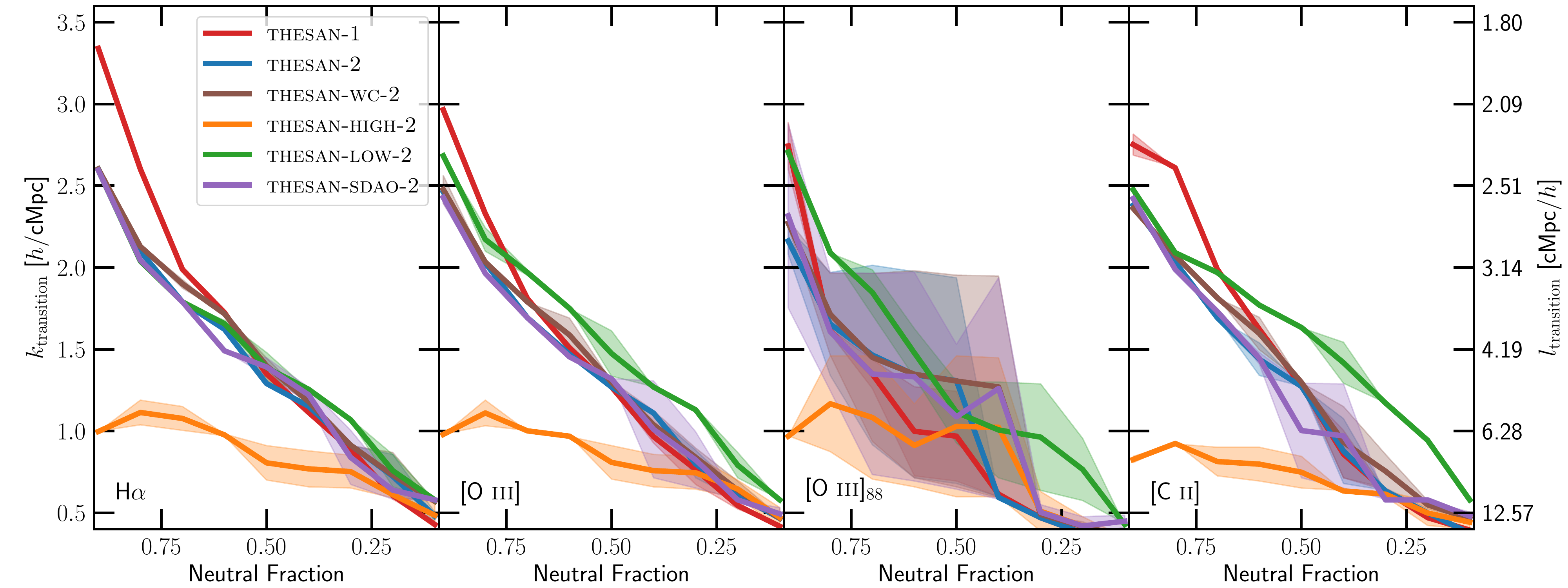}
    \caption{The transition wavenumber (\kt; left axis) and the corresponding transition scale ($l_\mathrm{transition}$; right axis) as a function of the volume-weighted neutral fraction in the Universe for the various emission lines and reionization models as indicated. While the general picture of \kt decreasing as reionization progresses is valid, its exact value and evolution depends substantially on the reionization model and to a lesser extent on the emission line. The simulations in which high mass haloes dominate the ionizing photon budget consistently show lower values of \kt and vice versa.}
    \label{fig:crossVal}
\end{figure*}

\subsection{Cross-correlation spectra}
In Figure~\ref{fig:cross_Ha_evolve} we show the cross-correlation PS between the 21\,cm emission and \halpha emission in \thesanone. The cross-correlation PS between two emission line intensity distributions $I_1(x)$ and $I_2(x)$ is defined as
\begin{equation}
    \Delta^2_{I_1-I_2}(k) = \frac{k^3}{2\pi^2} \frac{\left< \mathrm{Re}(\tilde{I}_1) \mathrm{Re}(\tilde{I}_2) + \mathrm{Im}(\tilde{I}_1) \mathrm{Im}(\tilde{I}_2) \right>}{V_\mathrm{box}}~,
\end{equation}
where $\tilde{I}(k)$ is the Fourier transform of $I(x)$ and Re$(\tilde{I})$ and Im$(\tilde{I})$ denote the real and Imaginary parts of $\tilde{I}(k)$ respectively. The colours indicate the volume weighted ionization fractions  ranging from $0.1$--$0.9$, in increments of $0.1$. A positive correlation between the quantities is indicated by a dashed curve while the solid curves signal a negative correlation. In the initial stages of the reionization process ($x_\ion{H}{II} \lesssim 0.1$), both the small and the large scale fluctuations correlate positively. The positive correlation on small scales can be explained by the fact that both the \halpha emission and the neutral hydrogen gas exists in the high density filaments and knots that host galaxies. On the largest scales, both emission line maps follow the underlying matter density distribution. They are anti-correlated on intermediate scales ($0.9 \lesssim k\,\mathrm{[}h\mathrm{/cMpc}\mathrm{]} \lesssim 4$), reflecting the relatively small sizes of these early \hii regions. By $x_\ion{H}{II} \sim 0.2$, the bubbles grow large enough that all the large scale modes ($k\,\mathrm{[}h\mathrm{/cMpc}\mathrm{]} \lesssim 3$) are anti-correlated. We note that even smaller wavenumbers might still be positively correlated, but the simulated volume is not large enough to capture these scales. As the bubble sizes increase, the point at which the cross-correlation PS switches sign, known as the transition wavenumber ($k_\mathrm{transition}$), moves to larger and larger spatial scales. Eventually all the spatial scales will show a positive correlation because the only neutral gas in the Universe will be present in the high density regions that also host galaxies that are the sites for \halpha emission.  $k_\mathrm{transition}$ can therefore be used to inform us about the progress of reionization and the neutral fraction evolution \citep{Gong2012}.

Of course, the exact value of the transition wavenumber will depend on the topology of the ionized bubbles and to a lesser extent on the underlying nebular emission line distribution. Therefore, the value of \kt and its evolution with redshift will allow us to place constraints on the sources responsible for the reionization process \citep{Dumitru2019}. In Figure~\ref{fig:crossAll} we plot the cross-correlation functions at volume weighted neutral fractions of $0.8$, $0.5$ and $0.2$, between 21\,cm and \halpha, \oiii, \oiii$_{88}$ and \cii as indicated. The different coloured lines indicate the different simulations in our suite, namely, \thesanone (red curves), \thesantwo (blue curves), \thesanwc (brown curves), \thesanhigh (orange curves), \thesanlow (green curves) and \thesansdao (purple curves). All reionization models show a relatively large \kt at low ionization fractions and decrease as the bubble sizes increase.  However, the exact value \kt and its evolution with the neutral fraction evolution is model dependent. Since \thesanhigh has the largest bubble sizes for a given ionization fraction, the transition happens at smaller wavenumbers when compared to other models. \thesanlow, on the other hand, generally shows the largest \kt, except at very low ionization fractions, where \thesanone shows the largest \kt because haloes at the very low mass end contribute the most during the initial stages of reionization. These haloes do not contribute in the other simulations, because they are not well resolved, thereby artificially reducing their star-formation rates. 

In Figure~\ref{fig:crossVal} we further quantify this behaviour by plotting \kt as a function of the volume weighted neutral fraction in the different simulations and for the different lines, as indicated. The shaded regions denote the uncertainty in \kt arising from the fact that the transition is not sharp because the  cross-correlation function fluctuates between positive and negative values before switching sign permanently. \kt is calculated as the mean of the smallest value of $k$ where the function is positive and the largest value of $k$ where it is negative and the uncertainty is just the span between these values. If there is more than one transition point, then we quote the values for the largest $k$. The errors in \kt are generally quite small, except for the \oiii$_{88}$ line which is shot-noise dominated except for the largest scales. 

 \begin{figure*}
	\includegraphics[width=1.07\textwidth]{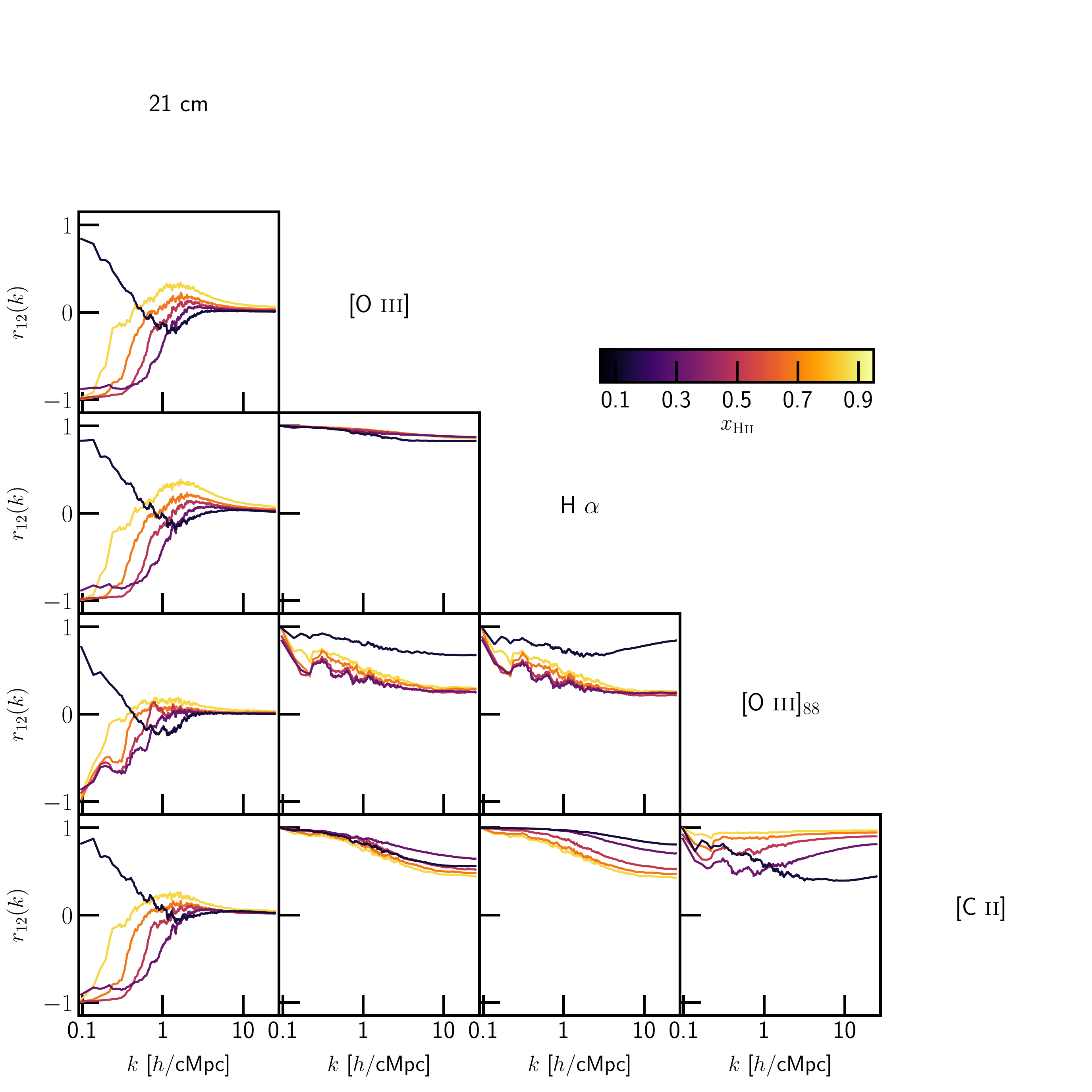}
    \caption{The cross-correlation coefficient ($r_{12}$;  defined in Eq.~\ref{eq:r12}) as a function of the wavenumber for the different emission lines considered in this work. The two intensity maps will be perfectly correlated or anti-correlated if $r_{12}(k)=1\,\mathrm{or}\,-1$, respectively. Values close to zero indicate de-correlation between the lines. \halpha, \oiii and \cii are well-correlated with each other on most spatial scales, while the steeper slope of the far IR line \oiii$_{88}$ makes it de-correlate rapidly at smaller $k$. }
    \label{fig:r12}
\end{figure*}

The general picture of \kt decreasing as reionization progresses is valid for all models and all the lines considered in this work. However, the exact value and its evolution depends heavily on the reionization model and to a lesser extent on the emission line. \thesanhigh shows the smallest \kt for all the lines considered at all neutral fractions, mirroring the fact that this reionization model has the highest contribution from large bubble sizes at any given ionization fraction \citep[see Section 3.4 of][]{KannanThesan}. The slope of the relation is quite shallow with \kt only changing by a factor of 2 over the majority of the reionization history. All the other simulations show a much steeper slope (with values changing by more than a factor of $5$), with \thesanone showing the steepest. This is because the initial stages of reionization are dominated by the very low mass haloes ($10^8$--$10^9$\,M$_\odot$) that are essentially unresolved in the other, lower resolution simulations. \thesanlow generally shows higher \kt when compared to other models because low bubble sizes dominate the reionization process. While this general picture is true for all emission lines considered in this work, there are slight differences in the exact value of \kt that might help place constraints on the slope of the \lsfr relation for the different lines.\footnote{See Appendix~\ref{app:ktslope} for a more thorough discussion on the dependence of \kt on the slope of the \lsfr relation.} It is therefore quite clear that the value of the transition wavenumber and its evolution provide important information that can help place stringent constraints on the bubble size distribution, which in turn informs us about the sources that dominate the ionizing photon budget at any particular redshift.

Finally, in Figure~\ref{fig:r12} we show the scale dependant correlation coefficient, $r_{12} (k)$, in \thesanone between 21\,cm and different nebular emission lines considered in this work for $x_\ion{H}{II}=0.1$--$0.9$ in increments of $0.2$ as indicated. This quantifies the amount of correlation between the two intensity maps and is defined as
\begin{equation}
    r_{12}(k) = \frac{P_{12}(k)}{\sqrt{P_{11}(k)P_{22}(k)}}~,
    \label{eq:r12}
\end{equation}
where $P_{12}(k)$ is the cross-correlation PS and $P_{11}(k)$ and $P_{22}(k)$ are the corresponding auto-correlation spectra. The two intensity maps will be perfectly correlated or anti-correlated if $r_{12}(k)=1\,\mathrm{or}\,-1$, respectively. Values close to zero indicate de-correlation between the lines. In general, the closer the slopes of the \lsfr relation the more correlated the nebular emission lines will be. It is important to quantify the cross-correlation coefficient because the auto-spectrum can be estimated from cross-spectra \citep{Beane2019}. This technique is more robust to residual foregrounds than the usual 21\,cm auto-power spectrum measurements and can help in verifying auto-spectrum detections. In fact, the larger the number of emission lines used the better the estimate of the spectrum. Importantly, $r_{12}$ also determines the ability to use one line as a tracer of the other, in order to subtract foregrounds/interlopers \citep{Schaan2021}, with values near one indicating good subtraction while values near zero indicate ineffective subtraction. Moreover, since the lines considered in this paper all trace slightly different gas phases in the ISM of star-forming galaxies, the correlated abundance of these lines can help constrain the ISM of high-redshift galaxies \citep{Sun2019}. 



Mirroring Figure~\ref{fig:crossAll}, cross-correlation coefficients between the 21\,cm emission and nebular emission lines show positive values at large scales in the initial stages of reionization. The intermediate wavenumbers are negatively correlated and the smallest scales are decorrelated. This picture changes rapidly, with most of the large spatial scales anti-correlating with the nebular emission lines by $x_\ion{H}{II}\gtrsim0.3$. The optical emission lines are well correlated ($r_{12}>0.8$) with each other on most spatial scales. However, the steeper slope of the far IR line \oiii$_{88}$ means that it only correlates with the optical lines on the largest scales and $r_{12}$ decreases steeply with $k$. The larger the difference between the slopes of the two luminosity relations, the more quickly the lines de-correlate. Since the slope of the \cii line matches the slope of the optical lines, it correlates well with them. Interestingly, the changing slope with redshift causes the line to de-correlate more with decreasing redshift (increased ionized fraction), which suggests that there is complementary information to be obtained by joint analyses of multiple emission lines.

\section{Discussion and Conclusions}
\label{sec:conclusions}
In this work we have presented a self-consistent framework to predict the intensity maps of a variety of nebular emission lines emanating from reionization era galaxies using the state-of-the-art \thesan simulations. The spectral energy distributions including the nebular emission line properties of the simulated galaxies are derived by interpolating between a grid of photo-ionization calculations using the \textsc{cloudy} code as described in \citet{Byler2017}. Dust attenuation is modelled using the Monte Carlo radiative transfer code \skirt \citep[last described in][]{skirt}, with the dust mass in each cell determined by a spatially constant dust to metal ratio (DTM) that was derived to match the observed UV luminosity function evolution from $z=2$--$10$ \citep{Vogelsberger2020}. The main results and conclusions of this work are as follows:
\begin{enumerate}
    \item We construct dust-attenuated SEDs for all well-resolved galaxies in the \thesanone simulation. The derived UV magnitudes match the observationally inferred luminosity function over a wide range of magnitudes. They are also in broad agreement with the $M_\star$--$M_\mathrm{UV}$ relation and measured UV continuum slope ($\beta_\mathrm{UV}$) of high-redshift galaxies. 
    
    \item We provide fitting functions for the luminosity--star-formation rate relations (\lsfr) of the brightest four rest frame optical emission lines, \mbox{\oii ($3726, 3729$\,\AA)}, \mbox{\hbeta ($4861$\,\AA)}, \mbox{\oiii ($4959, 5007$\,\AA)} and \mbox{\halpha ($6563$\,\AA)}, and four fine structure emission lines in the far IR, \mbox{\oiii$_{52}$ ($52$\,\micron)}, \mbox{\oiii$_{88}$ ($88$\,\micron)}, \mbox{\nii ($122$\,\micron)} and \mbox{\cii ($158$\,\micron)}.
    
    \item We determine that important differences exist between the derived \lsfr relation for high-$z$ galaxies and the commonly used low-$z$ estimates. Specifically, the hydrogen emission lines show shallower scaling relations and slightly higher amplitudes, while the metal lines show steeper slopes with generally lower amplitudes. These differences can be attributed to the fact that high-$z$ galaxies are generally less metal enriched than their low-$z$ counterparts.
    
    \item Our predicted mean intensities of the Balmer emission lines are about a factor of $3-4$ larger than previous estimates \citep[see for example,][]{Gong2017}. This disparity is mostly due to the higher LyC production rate in low metallicity stars efficiently forming within lower-mass haloes, which was not accounted for in previous works.
    
    \item We show that the cross-correlation power spectra between the 21\,cm emission and various nebular emission lines can be used to infer the evolution of the neutral hydrogen fraction during reionization. The scale at which the cross-correlation PS switches sign, knows as the transition wavenumber (\kt), moves to larger spatial scales as reionization progresses.
    
    \item The value of \kt and its evolution depend strongly on the reionization model and to a lesser extent on the emission line considered. Scenarios in which massive haloes dominate the ionizing photon budget show consistently lower values of \kt and vice-versa. This is consistent with the picture that the transition scale probes the typical sizes of ionized regions.
    
    \item The cross-correlation coefficients show that the nebular emission lines \halpha, \oiii and \cii are well-correlated with each other on most spatial scales. However, the steeper slope of the far IR line \oiii$_{88}$ means that it only correlates with the other lines on the largest scales and de-correlates rapidly at smaller spatial scales, where shot noise dominates. 
\end{enumerate}

The emission line modelling presented in this work does a relatively good job of estimating the luminosities of the lines arising from highly ionized \hii regions around newly formed young stars. However, it is unable to predict the luminosities of lines which predominantly originate in PDRs such as \cii ($158$\,\micron) and \oi ($63$\,\micron). Accurate modelling of these lines will require additional assumptions about the prevalence and distribution of giant molecular clouds and the spatial distribution of the far-UV radiation field \citep{Olsen2015,Olsen2017}. In  a similar vein, line luminosities of the various CO lines would require assumptions about the molecular gas fractions within galaxies. While scaling relations like the ones presented in \citet{Leroy2008} and \citet{GD2014} can be used to estimate the molecular gas content, they usually rely on the physics of low redshift, high metallicity galaxies, where the main \ion{H}{$_2$} formation channel is via dust grains. This approximation is unlikely to be accurate in low metallicity environments where a series of gas phase reactions involving intermediary ions like \ion{H}{$^-$} and \ion{H}{$^+_2$} are the primary channels for molecular gas formation \citep{Glover2005}. Thus, self-consistent CO LIM predictions would ideally require more rigorous modelling of the multi-phase ISM coupled with molecular thermochemistry and radiation field intensities \citep{Katz2019, Kannan2020b}, which we leave for future work. 

Beyond this, the prominent \lya transition ($2$--$1$) of neutral hydrogen is an exceptionally promising target for upcoming intensity mapping experiments like \texttt{SPHEREx} and \texttt{CDIM} \citep[e.g.][]{Visbal2018,Mas-Ribas2020}. Part of the utility of \lya stems from the additional physics of resonant absorption, which effectively removes \lya photons out of the line of sight to be lost to the diffuse cosmic optical and infrared backgrounds \citep{GunnPeterson1965}. This encodes information about both the sources (galaxies) and sinks (IGM) for a powerful probe of reionization, e.g. \lya emitting galaxies (LAEs) and the \lya forest in emission and absorption, respectively \citep[in the context of \thesan, see][]{GaraldiThesan,SmithThesan}. However, in practice the theoretical modelling and observational interpretations are both significantly more complex because the results are highly sensitive to the radiative transfer within galaxies and through the intervening IGM \citep{Dijkstra2019,Ouchi2020}. In particular, the emergent \lya spectral flux from high-$z$ galaxies is very sensitive to neutral hydrogen and dust column densities, feedback regulated outflows and morphologies, and time and anisotropy variability, all of which increase the uncertainty of modelling \lya properties during the Epoch of Reionization with important consequences for \lya LIM as well \citep[e.g.][]{Smith2017,Smith2019,Behrens2019,Laursen2019,Garel2021}. In a future accompanying paper (Smith et al. in prep.) we will provide \lya intensity mapping predictions based on high time cadence on-the-fly redshift-space renderings of \lya properties. In combination with insights from detailed Monte Carlo \lya radiative transfer simulations and IGM transmission studies, we hope to similarly advance our understanding of \lya LIM theory, modelling, and interpretation.

Multi-tracer line intensity mapping, by nature, requires a variety of observational instruments to measure lines at different wavelengths, each with its own sensitivity and instrument-specific noise power spectrum. Moreover, the signal-to-noise ratios depend on cost optimisations such as beam sizes and integration times. Therefore, placing constraints on any astrophysical and cosmological parameters would require an intimate knowledge of these instrumental effects, survey strategies, and realistic signal contaminants. We therefore choose to defer predictions for the detectability of the various lines and the constraining power of these measurements to a future paper. We hope that the emission line scaling relations and predictions for the line intensity maps derived in this work will provide a rigorous theoretical framework upon which the results from current and upcoming LIM experiments can be analysed and interpreted. 

\section*{Acknowledgements}

We thank the anonymous referee for constructive comments that helped improve the paper. We also thank Angus Beane for insightful discussions related to this work.
AS acknowledges support for Program number \textit{HST}-HF2-51421.001-A provided by NASA through a grant from the Space Telescope Science Institute, which is operated by the Association of Universities for Research in Astronomy, incorporated, under NASA contract NAS5-26555. MV acknowledges support through NASA ATP grants 16-ATP16-0167, 19-ATP19-0019, 19-ATP19-0020, 19-ATP19-0167, and NSF grants AST-1814053, AST-1814259,  AST-1909831 and AST-2007355. The authors gratefully acknowledge the Gauss Centre for Supercomputing e.V. (\url{www.gauss-centre.eu}) for funding this project by providing computing time on the GCS Supercomputer SuperMUC-NG at Leibniz Supercomputing Centre (\url{www.lrz.de}). Additional computing resources were provided by the Extreme Science and Engineering Discovery Environment (XSEDE), at Stampede2 and Comet through allocation TG-AST200007  and by the NASA High-End Computing (HEC) Program through the NASA Advanced Supercomputing (NAS) Division at Ames Research Center.

\section*{Data Availability}

 All simulation data, including intrinsic and dust attenuated galaxy SEDs will be made publicly available in the near future. Data will be distributed via \url{www.thesan-project.com}. Before the public data release, data underlying this article will be shared on reasonable request to the corresponding author(s).



\bibliographystyle{mnras}
\bibliography{references} 




\appendix
\section{Extrapolation convergence}
\label{app:extrp}
 \begin{figure}
	\includegraphics[width=0.99\columnwidth]{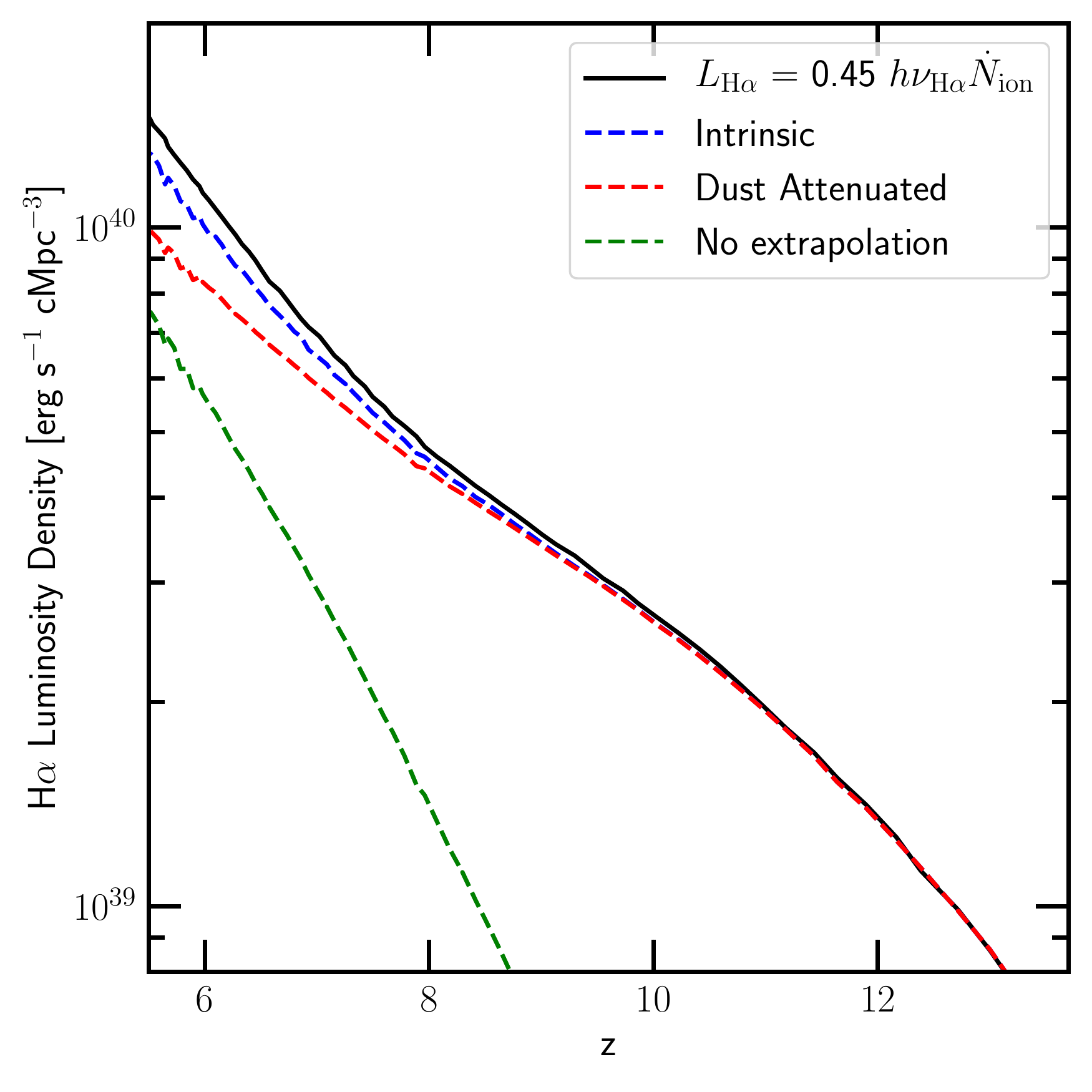}
    \caption{The theoretically calculated \halpha luminosity density as a function of redshift (black curve) compared to the \halpha luminosity values obtained from using the \lsfr fitting relation derived from intrinsic (blue curve) and dust-attenuated SEDs (red curve). For comparison the green curve plots the \halpha luminosity if only the values from well-resolved haloes (as defined in Section~\ref{sec:methods}) are considered. The fact that the extrapolated \halpha luminosities are close to the theoretically expected values, especially at high redshifts, shows the goodness of the fits and the effectiveness of extrapolating the fits to all haloes in the simulation.}
    \label{fig:extrp}
\end{figure}
One of the major assumptions in our model is that the \lsfr fits derived from the emission line calculations of the well-resolved galaxies (as defined in Section~\ref{sec:methods}) are also valid for the lower-mass galaxies in the simulation volume. We test the validity of this extrapolation by comparing the \halpha luminosity density derived from the \lsfr fits applied to all galaxies in the simulation volume to the theoretically expected value as shown in Figure~\ref{fig:extrp}. The theoretical value is derived by assuming that all LyC photons in the simulation lead to the ionization of a hydrogen atom. These atoms recombine leading to \halpha line emission with a conversion probability per recombination event of $0.45$ (black curve). The total LyC production rate is calculated by summing up the ionizing emissivity of all the stars in the simulation volume. For comparison, the plot also shows the \halpha luminosity density obtained from using the \lsfr fitting relation derived from intrinsic (defined as the sinlge power law relation with slope $m_a$ and intercept $a$ as described by Eq.~\ref{eq:fit}; blue curve) and dust-attenuated SEDs (red curve). Finally, the green curve plots the luminosity if only the well-resolved haloes are considered. The fact that the extrapolated \halpha luminosities are close to the theoretically expected values, especially at high redshifts, shows the reliability of the fits and the effectiveness of extrapolating the fits to all haloes in the simulation. We note that the decrease in the \halpha luminosity at low-$z$ is due to dust attenuation, which becomes increasingly important at lower redshifts. The discrepancy between the theoretical and intrinsic fits arises from the fact that at low-$z$, the ISM becomes more metal enriched, meaning that a larger fraction of the ionizing photons is used up for ionizing metals rather than hydrogen.

\section{Dependence of \kt on the slope of the \lsfr relation}
\label{app:ktslope}
\begin{figure}
	\includegraphics[width=0.99\columnwidth]{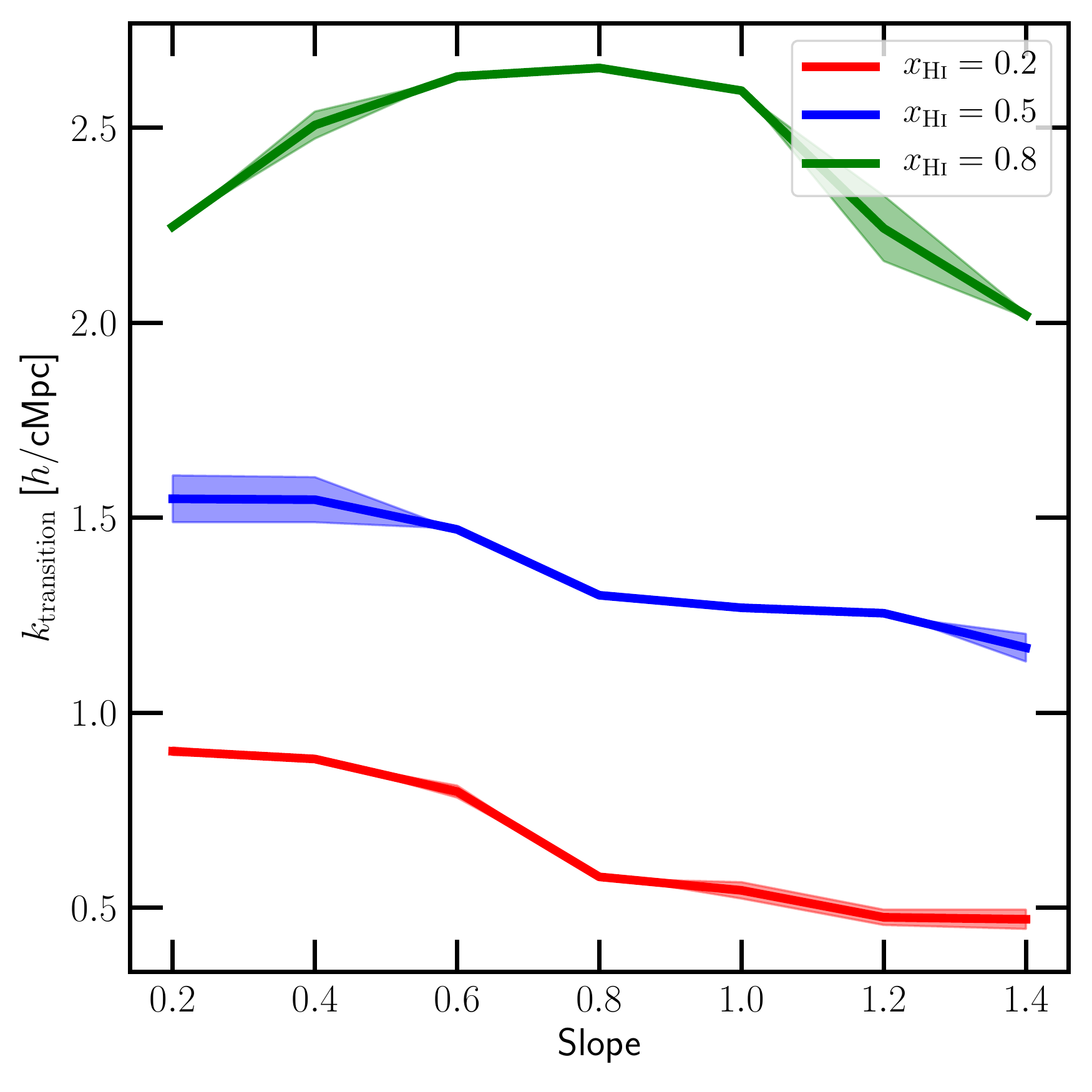}
    \caption{The value of \kt as a function of the slope of the \lsfr relation at three different neutral fractions of $x_\ion{H}{I} = 0.2$ (red curve), $x_\ion{H}{I} = 0.5$ (blue curve), and $x_\ion{H}{I} = 0.8$ (green curve) in the \thesanone simulation. While \kt generally traces the reionization history, there is a non-negligible dependence on the slope of the \lsfr relation. It is therefore important to accurately model the \lsfr relation in order to reliably constrain the reionization history of the Universe using cross-correlation spectra. }
    \label{fig:ktslope}
\end{figure}
The value of \kt is usually assumed to trace the average scale of the ionized bubble during the reionization process \citep{Dumitru2019}, with  the impact of the slope of the \lsfr relation assumed to be minimal. Here we test this hypothesis by changing the slope of \lsfr relation from $0.2$--$1.4$ and cross-correlating with the 21\,cm filed at three different values for the ionization fractions as indicated in Figure~\ref{fig:ktslope}.  We note that for slopes of $m\gtrsim1$, the highest luminosity galaxies start to dominate the total emission luminosity in the line, reducing the significance of the fluctuations arising from lower luminosity haloes. This leads to an increase in shot noise contribution even at relatively larger spatial scales. We therefore, choose to cap the slope at the high SFR end ($x_c$ in Eq.~\ref{eq:fit}) to $0.8$. This is equivalent to masking out the most luminous haloes in the volume so as to highlight the lower luminosity features in the LIM.  As expected \kt shows a strong dependence on the ionization fraction, in agreement with the picture that it traces the average size of the ionized regions. Nonetheless,  there is a non-negligible dependence on the slope of the \lsfr relation. It is therefore important to accurately model the \lsfr relation in order to reliably constrain the reionization history of the Universe using cross-correlation spectra. However, we do note that the slope of the \lsfr relation has to be really poorly constrained in order for it to produce an appreciable uncertainty in \kt. If the slope is constrained within a factor of $0.2$, then the uncertainty in \kt is pretty small, and this quantity can therefore be used as a rigorous probe of the reionization process. 






\bsp	
\label{lastpage}
\end{document}